\pdfoutput=1

\PassOptionsToPackage{hyphens}{url}
\documentclass[format=sigconf,screen=true]{acmart}
\usepackage{etoolbox}
\usepackage{graphicx}
\usepackage{balance}

\usepackage[ruled]{algorithm2e}
\SetKwProg{Fn}{function}{}{end}
\SetKwProg{Macro}{macro}{}{end}
\SetKwProg{Init}{initialization}{}{}
\SetKwInput{Input}{input}
\SetKw{KwTo}{in}
\SetKw{KwWhere}{where}

\usepackage{setspace}
\usepackage{listings}
\usepackage{multicol}
\usepackage{xspace}
\usepackage{enumitem}
\usepackage{booktabs}
\usepackage{siunitx}
\usepackage{color}

\newcommand{\minihead}[1]{{\vspace{.45em}\noindent\textbf{#1.} }}

\newcommand*{\msketch}{moments sketch\xspace}
\newcommand*{\msketches}{moments sketches\xspace}
\newcommand*{\racz}{RTTBound\xspace}
\newcommand*{\markov}{MarkovBound\xspace}
\newcommand*{\stree}{\texttt{S-Hist}\xspace}
\newcommand*{\strees}{\texttt{S-Hists}\xspace}

\newcommand*{\yahoo}{\texttt{Merge12}\xspace}
\newcommand*{\tdigest}{\texttt{T-Digest}\xspace}
\newcommand*{\gk}{\texttt{GK}\xspace}
\newcommand*{\sampling}{\texttt{Sampling}\xspace}

\newcommand*{\hist}{\texttt{EW-Hist}\xspace}
\newcommand*{\random}{\texttt{RandomW}\xspace}

\newtoggle{rcolors}
\newcommand{\colora}[1]{\iftoggle{rcolors}{{\color{red}{#1}}}{#1}}

\toggletrue{rcolors}
\togglefalse{rcolors}

\newtoggle{arxiv}
\toggletrue{arxiv}

\iftoggle{arxiv}{
    \settopmatter{printacmref=false, printccs=false, printfolios=true}
    \renewcommand\footnotetextcopyrightpermission[1]{} 
    \pagestyle{plain} 
    \setcopyright{none} 
}{
    \usepackage{cite}
    \usepackage[hyphens]{url}
    \vldbTitle{Moment-Based Quantile Sketches for Efficient High Cardinality Aggregation Queries}
    \vldbAuthors{Edward Gan, Jialin Ding, Kai Sheng Tai, Vatsal Sharan, and Peter Bailis}
    \vldbDOI{https://doi.org/TBD}
}

\begin{document}

\iftoggle{arxiv}{
\title{Moment-Based Quantile Sketches\\for Efficient High Cardinality Aggregation Queries}
\titlenote{A version of this paper has been accepted to VLDB 2018. This document is its associated technical report.}
\author{Edward Gan, Jialin Ding, Kai Sheng Tai, Vatsal Sharan, Peter Bailis}
\affiliation{
    \institution{Stanford InfoLab}
}
\begin{abstract}
Interactive analytics increasingly involves querying for quantiles over sub-populations of high cardinality datasets.
Data processing engines such as Druid and Spark use mergeable summaries to estimate quantiles, but summary merge times can be a bottleneck during aggregation.
We show how a compact and efficiently mergeable quantile sketch can support aggregation workloads. 
This data structure, which we refer to as the \msketch, operates with a small memory footprint (200 bytes) and computationally efficient (50ns) merges by tracking only a set of summary statistics, notably the sample moments.
We demonstrate how we can efficiently estimate quantiles using the method of moments and the maximum entropy principle, and show how the use of a cascade further improves query time for threshold predicates.
Empirical evaluation shows that the moments sketch can achieve less than 1 percent quantile error with $15\times$ less overhead than comparable summaries, improving end query time in the MacroBase engine by up to $7\times$ and the Druid engine by up to $60\times$.
\end{abstract}
\maketitle
}{
\title{Moment-Based Quantile Sketches\\for Efficient High Cardinality Aggregation Queries}
\author{
    Edward Gan, Jialin Ding, Kai Sheng Tai, Vatsal Sharan, Peter Bailis\\
    \affaddr{Stanford InfoLab}
}
\maketitle
\begin{abstract}
Interactive analytics increasingly involves querying for quantiles over sub-populations of high cardinality datasets.
Data processing engines such as Druid and Spark use mergeable summaries to estimate quantiles, but summary merge times can be a bottleneck during aggregation.
We show how a compact and efficiently mergeable quantile sketch can support aggregation workloads. 
This data structure, which we refer to as the \msketch, operates with a small memory footprint (200 bytes) and computationally efficient (50ns) merges by tracking only a set of summary statistics, notably the sample moments.
We demonstrate how we can efficiently estimate quantiles using the method of moments and the maximum entropy principle, and show how the use of a cascade further improves query time for threshold predicates.
Empirical evaluation shows that the moments sketch can achieve less than 1 percent quantile error with $15\times$ less overhead than comparable summaries, improving end query time in the MacroBase engine by up to $7\times$ and the Druid engine by up to $60\times$.
\end{abstract}
}

\section{Introduction}
\label{sec:intro}
Performing interactive multi-dimensional analytics over data from sensors, devices, and servers increasingly requires computing aggregate statistics for specific subpopulations and time windows~\cite{rabkin2014agg,feng2015streamcube,agarwal2013blinkdb}.
In applications such as A/B testing~\cite{hill2017bandit,johari2017peeking}, exploratory data analysis~\cite{bailis2017macrobase,vartak2015seedb}, and operations monitoring~\cite{beyer2016site,Abraham2013scuba}, analysts perform aggregation queries to understand how specific user cohorts, device types, and feature flags are behaving.
In particular, computing quantiles over these subpopulations is an essential part of debugging and real-time monitoring workflows~\cite{dean2013tail}.

As an example of this quantile-driven analysis, our collaborators on a Microsoft application monitoring team collect billions of telemetry events daily from millions of heterogeneous mobile devices.
Each device tracks multiple metrics including request latency and memory usage, and is associated with dimensional metadata such as application version and hardware model.
Engineers issue quantile queries on a Druid-like~\cite{Yang2014druid} in-memory data store, aggregating across different dimensions to monitor their application (e.g., examine memory trends across device types) and debug regressions (e.g., examine tail latencies across versions).
Querying for a single percentile in this deployment can require aggregating hundreds of thousands of dimension value combinations.

\begin{figure}
\includegraphics[width=\columnwidth]{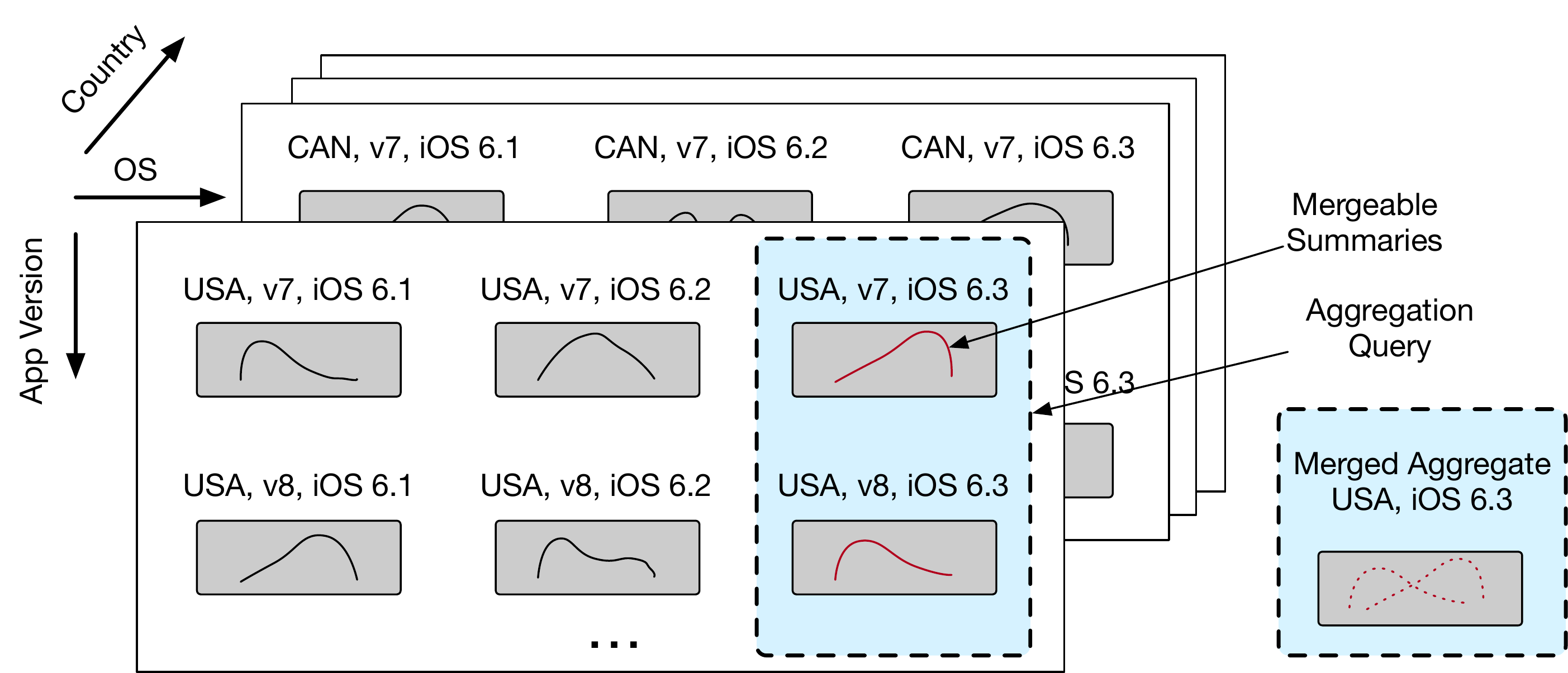}
\caption{Given a data cube with pre-aggregated summaries, we can compute roll-ups by merging the relevant summaries. Efficiently mergeable summaries enable scalable aggregations.}
\vspace{-1em}
\label{fig:aggregation}
\end{figure}
When users wish to examine the quantiles of specific slices of a dataset, OLAP engines such as Druid and Spark support computing approximate quantiles using compressed representations (\emph{summaries}) of the data values \cite{druidblog,Yang2014druid,sparkblog,zaharia2012resilient}.
By pre-aggregating a summary for each combination of dimension values, Druid and similar engines can reduce query times and memory usage by operating over the relevant summaries directly, effectively constructing a data cube~\cite{gray1997datacube,rabkin2014agg}.

\colora{
Given a time interval and a metric with $d$ associated dimensions, Druid maintains one pre-aggregated summary for each $d$-tuple of dimension values.
These summaries are kept in RAM across a number of nodes, with each node scanning relevant summaries to process subsets of the data specified by a user query.
Figure~\ref{fig:aggregation} illustrates how these \emph{mergeable}~\cite{agarwal2012mergeable} summaries can be aggregated to compute quantile roll-ups along different dimensions without scanning over the raw data.

More concretely, a Druid-like data cube in our Microsoft deployment with 6 dimension columns, each with $10$ distinct values, is stored as a set of up to $10^6$ summaries per time interval.
On this cube, computing the 99-th percentile latency for a specific app version can require 100,000 merges, or even more for aggregation across complex time ranges.
When there are a limited number of dimensions but enormous data volumes, it is cheaper to maintain these summaries than scan over billions of raw datapoints.
}

Many quantile summaries support the merge operation~\cite{agarwal2012mergeable,greenwald2001space,tdigest}, but their runtime overheads can lead to severe performance penalties on high-cardinality datasets.
Based on our experiments (Section~\ref{sec:evalquerytime}), one million 1KB GK-sketches \cite{greenwald2001space} require more than 3 seconds to merge sequentially, limiting the types of queries users can ask interactively.
The merging can be parallelized, but additional worker threads still incur coordination and resource usage overheads. Materialized views \cite{liu2016kodiak,nandi2011distributed,kamat2014dice,Harinarayan1996Implementing}, sliding window sketches \cite{datar2002maintaining}, and dyadic intervals can also reduce this overhead.
However, dyadic intervals only apply to ordered dimensions and maintaining materialized views for multiple dimension roll-ups can be prohibitively expensive in a real-time stream, so merge time remains a relevant bottleneck.

In this paper, we enable interactive quantile queries over high-cardinality aggregates by introducing a compact and efficiently mergeable quantile sketch and associated quantile estimation routines.
We draw a connection between the classic \emph{method of moments} for parameter estimation in statistics~\cite{wasserman2010all} and the need for efficient summary data structures.
We show that storing the sample moments $\mu_i = \frac{1}{n}\sum x^i$ and log-moments $\nu_i = \frac{1}{n}\sum \log^i{(x)}$ can enable accurate quantile estimation over a range of real-world datasets while utilizing fewer than 200 bytes of memory and incurring merge times of less than 50 nanoseconds.
In the context of quantile estimation, we refer to our proposed summary data structure as the \emph{\msketch}.

While constructing the \msketch is straightforward, the inverse problem of estimating quantiles from the summary is more complex.
The statistics in a \msketch provide only loose constraints on the distribution of values in the original dataset: many distributions might match the moments of a \msketch but fail to capture the dataset structure.
Therefore, we make use of the \emph{principle of maximum entropy}~\cite{jaynes1957} to compute a ``least-biased'' quantile estimate for a \msketch.
On continuous real-valued datasets, we find that this approach yields more accurate estimates than alternative methods, achieving $\epsilon \leq 1\%$ error with 200 bytes of memory.
To achieve this, we also describe a series of practical optimizations to standard entropy maximization that allow us to compute quantile estimates in under 1 millisecond on a range of real-world datasets. 

These query times make the \msketch a suitable summary when many merges (hundreds of thousands) are required, memory per-summary may be limited to less than 1 kilobyte, and $\epsilon = .01$ error is acceptable.
The \msketch and our maximum entropy estimate is most useful in datasets without strong discretization and when very small $< 10^{-4}$ error is not required.
The maximum entropy principle is less accurate when there are clusters of discrete values in a dataset (Section~\ref{sec:evalacc}), and floating point stability (Section~\ref{sec:choosek}) limits the minimum achievable error using this approach.

Moving beyond simple quantile queries, many complex queries depend on the quantile estimates of multiple subpopulations.
For example, data exploration systems such as MacroBase~\cite{bailis2017macrobase} are interested in finding all subpopulations that match a given threshold condition (e.g., subpopulations where the 95th percentile latency is greater than the global 99th percentile latency).
Given a large number of subpopulations, the cost of millisecond-level quantile estimates on thousands of subgroups will accumulate.
Therefore, to support threshold queries over multiple populations, we extend our quantile estimator with a \emph{cascade}~\cite{viola2001rapid}, or sequence of increasingly precise and increasingly expensive estimates based on bounds such as the Markov inequalities. 
For queries with threshold conditions, the cascades dramatically reduce the overhead of quantile estimation in a \msketch, by up to $25\times$.

We implement the \msketch both as a reusable library and as part of the Druid and MacroBase analytics engines.
We empirically compare its accuracy and efficiency with alternative mergeable quantile summaries on a variety of real-world datasets. 
We find that the \msketch offers $16$ to $50\times$ faster merge times than alternative summaries with comparable accuracy.
This enables $15$ to $50\times$ faster query times on real datasets.
Moreover, the \msketch enables up to $7\times$ faster analytics queries when integrated with MacroBase and $60\times$ faster end-to-end queries when integrated with Druid.

In summary, we make the following contributions:
\begin{itemize}
\setlength\itemsep{.5em}
\setlength{\topsep}{0em} 
\item We illustrate how statistical moments are useful as efficient mergeable quantile sketches in aggregation and threshold-based queries over high-cardinality data.
\item We demonstrate how statistical and numerical techniques allow us to solve for accurate quantile estimates in less than 1 ms, and show how the use of a cascade further improves estimation time on threshold queries by up to $25\times$.
\item We evaluate the use of moments as quantile summaries on a variety of real-world datasets and show that the \msketch enables $15$ to $50\times$ faster query times in isolation, up to $7\times$ faster queries when integrated with MacroBase and up to $60\times$ faster queries when integrated with Druid over comparably-accurate quantile summaries.
\end{itemize}

The remainder of this paper proceeds as follows.
In Section~\ref{sec:related}, we discuss related work.
In Section~\ref{sec:background}, we review relevant background material.
In Section~\ref{sec:algorithm}, we describe the proposed \msketch.
In Section~\ref{sec:alg_tresholds}, we describe a cascade-based approach for efficiently answering threshold-based queries.
In Section~\ref{sec:eval}, we evaluate the \msketch in a series of microbenchmarks.
In Section~\ref{sec:evalapps}, we evaluate the \msketch as part of the Druid and MacroBase systems, and benchmark its performance in a sliding window workflow.
We conclude in Section~\ref{sec:conclusion}.
\colora{
We include supplemental appendices in an extended technical report~\cite{gan2018momentext}.
}
\section{Related Work}
\label{sec:related}

\minihead{High-performance aggregation}
The aggregation scenarios in Section~\ref{sec:intro} are
found in many existing streaming data systems~\cite{Yang2014druid,rabkin2014agg,Braun2015analytics,cranor2003giga,bailis2017macrobase}, as well as data cube~\cite{sarawagi2000user,gray1997datacube}, data exploration~\cite{Abraham2013scuba}, and visualization~\cite{budiu2016sketch} systems.
In particular, these systems are can perform interactive aggregations over time windows and along many cube dimensions, motivating the design of our sketch. 
Many of these systems use approximate query processing, sampling, and summaries to improve query performance \cite{agarwal2013blinkdb,moritz2017trust,hall2016tradeoff}, but do not develop data structures specific to quantiles.
We believe the \msketch serves as a useful primitive in these engines.

Sensor networking is a rich source of algorithms for heavily resource-constrained settings. 
Sensor network aggregation systems \cite{Madden2002tag} support large scale roll-ups, but work in this area is largely focused on the complementary problem of communication plans over a network~\cite{kempe2003gossip,manjhi2005tributaries,cormode2007streaming}. 
Mean, min, max, and standard deviation in particular are used in \cite{Madden2002tag} as functions amenable to computation-constrained environments, but the authors do not consider higher moments or their application to quantile estimation. 

Several database systems make use of summary statistics in general-purpose analytics.
Muthukrishan et al~\cite{muthukrishnan2005streams} observe that the moments are a convenient statistic in streaming settings and use it to fill in missing integers.
Data Canopy~\cite{wasay2017canopy} uses first and second moments as an efficiently mergeable statistic for computing standard deviations and linear regressions. 
Similarly, systems on probabilistic data cubes such as~\cite{xie2016olap} use the first and second moments to prune queries over cube cells that store distributions of values.
In addition, many methods use compressed data representations to perform statistical analyses such as linear regression, logistic regression, and PCA~\cite{shanmugasundaram1999compressed,chen2006regression,xi2009compression,ordonez2016gamma}.
We are not aware of prior work utilizing higher moments to efficiently estimate quantiles for high-dimensional aggregation queries. 

\minihead{Quantile summaries}
There are a variety of summary data structures for the $\epsilon$-approximate quantile estimation problem \cite{buragohain2009quantiles,greenwald2001space,shrivastava2004medians,Cormode2005countmin}. 
Some of these summaries assume values from a fixed universe \cite{shrivastava2004medians,Cormode2005countmin}, while others operate using only comparisons \cite{greenwald2001space,agarwal2012mergeable}.
Our proposed \msketch and others~\cite{tdigest,ben2010streaming} operate on real values.
Agarwal et al.~\cite{agarwal2012mergeable} provide the initial motivation for mergeable summaries, as well as a proposed mergeable quantile sketch.
\colora{The authors in~\cite{luo2016quantiles,wang2013quantiles} benchmark a variety of quantile summaries but do not directly evaluate merge time.
Zhuang~\cite{zhuang2015experimental} evaluates merge performance of a variety of quantile summaries in a distributed setting, finding the \texttt{Random} summary to be the fastest.
To our knowledge we are the first to introduce and evaluate the \msketch for fast merge times and low space overhead.
}

\minihead{Method of moments} The method of moments is a well-established statistical technique for estimating the parameters of probability distributions~\cite{wasserman2010all}.
The main idea behind this approach is that the parameters of a distribution of interest $P$ can be related to the expectations of functions of the random variable $X\sim P$.
As a general method for consistent statistical parameter estimation, the method of moments is used across a wide range of fields, including econometrics~\cite{hansen1982large}, physics~\cite{mead1984maxent,gibson2014method}, and machine learning~\cite{belkin2010polynomial,kalai2010efficiently,anandkumar2012method}.
In this work, we demonstrate how the method of moments, applied in conjunction with practical performance optimizations, can scale to support real-world latency-sensitive query processing workloads.

\minihead{Maximum entropy principle}
The maximum entropy principle prescribes that one should select the least informative distribution that is consistent with the observed data. 
In the database community, this principle has been applied to estimating cardinality ~\cite{srivastava2006isolver} and predicate selectivity~\cite{markl2007consistent}.
Mead and Papanicolaou~\cite{mead1984maxent} apply the maximum entropy principle to the problem of estimating distributions subject to moment constraints; follow-up work proposes the use of Chebyshev polynomials for stability~\cite{silver1997calculation,badyo2005max} and faster approximation algorithms~\cite{balestrino2003maxent}, though we have not seen any practical implementations suitable for use in a database.
The maximum entropy principle is also used in machine learning, notably in the context of \emph{maximum entropy models}~\cite{berger1996maximum}.
For example, in natural language processing, maximum entropy models are a popular choice for tasks such as text classification~\cite{Nigam99usingmaximum} and sequence tagging~\cite{lafferty2001conditional}.

\section{Background}
\label{sec:background}
In this section, we review the approximate quantile estimation problem, mergeable quantile summaries, and our target query cost model.

\subsection{Quantile Queries}
\label{sec:quantilebg}
Given a dataset $D$ with $n$ elements, for any $\phi \in (0, 1)$, the \emph{$\phi$-quantile} of $D$ is the item $x\in D$ with rank $r(x) = \lfloor \phi n \rfloor$, where the rank of an element $x$ is the number of elements in $D$ smaller than $x$.

An $\epsilon$-approximate $\phi$-quantile is an element with rank between $(\phi - \epsilon) n$ and $(\phi + \epsilon) n$ \cite{agarwal2012mergeable}.
Given an estimated $\phi$-quantile $\hat{q}_{\phi}$, we can also define its \emph{quantile error} $\varepsilon$~\cite{luo2016quantiles} as the following:
\begin{equation}
\varepsilon = \frac{1}{n}\left|\text{rank}\left(\hat{q}_\phi\right) - \lfloor \phi n \rfloor\right|,
\label{eqn:quantile_error}
\end{equation}
such that an $\epsilon$-approximate quantile has error at most $\varepsilon$.
For example, given a dataset $D=\{1,\dots,1000\}$, an estimate $\hat{q}_{0.5} = 504$ for the $\phi=0.5$ quantile would have error $\varepsilon=0.003$.
In this paper, we consider datasets $D$ represented by collections of real numbers $D \subset \mathbb{R}$.

\emph{Quantile summaries} are data structures that provide approximate quantile estimates for a dataset given space sub-linear in $n$.
These summaries usually have a parameter $k_\epsilon$ that trades off between the size of the summary and the accuracy of its estimates. 
An $\epsilon$-approximate quantile summary provides $\epsilon$ approximate $\phi$-quantiles, where $\epsilon$ can be a function of space usage and the dataset \cite{buragohain2009quantiles,greenwald2001space,shrivastava2004medians,Cormode2005countmin}.

\subsection{Mergeable Summaries}
\label{sec:summarybg}
Agarwal et al.~\cite{agarwal2012mergeable} introduce the concept of \emph{mergeability} to accurately combine summaries in distributed settings.
Formally, for a summary with parameter $k_\epsilon$, we use $S(D, k_\epsilon)$ to denote a valid summary for a dataset $D$.
For any pair of datasets $D_1$ and $D_2$, the summarization routine $S$ is \emph{mergeable} if there exists an algorithm (i.e., the ``merge'' procedure) that produces a combined summary 
\begin{equation*}
    S(D_1 \uplus D_2, k_\epsilon) = \text{merge}(S(D_1, k_\epsilon), S(D_2, k_\epsilon))
\end{equation*} 
from any two input summaries, where $\uplus$ denotes multiset addition. 

Intuitively, a summary is mergeable if there is no accuracy cost to combining pre-computed summaries compared with constructing a summary on the raw data.
Thus, mergeable summaries are \emph{algebraic} aggregate functions in the data cube literature~\cite{gray1997datacube}.
As an example, an equi-depth histogram \cite{cormode2012synopses} on its own is not mergeable because there is no way to accurately combine two overlapping histogram buckets without access to additional data. 

Mergeable summaries can be incorporated naturally into a variety of distributed systems. 
In the MapReduce paradigm, a ``map'' function can construct summaries over shards while a ``reduce'' function merges them to summarize a complete dataset \cite{agarwal2012mergeable}.
\colora{In the GLADE system~\cite{rusu2012glade}, mergeable summaries are an example of a \emph{Generalized Linear Aggregate} (GLA), a user-defined computation that can be incrementally aggregated across nodes.}

\subsection{Query Model}
\label{sec:costmodel}
As described in Section~\ref{sec:intro}, we focus on improving the performance of quantile queries over aggregations on high cardinality datasets.
Given a dataset with $d$ categorical dimensions, we consider data cubes that maintain summaries for every $d$-way dimension value tuple as one natural setting for high performance aggregations, and many other settings are also applicable~\cite{wasay2017canopy}.
In these settings, query time is heavily dependent on the number of merges and the time per merge.

We consider two broad classes of queries in this paper.
First, \emph{single quantile} queries ask for quantile estimates for a single specified population.
For example, we can query the 99th percentile of latency over the last two weeks for a given version of the application:
\begin{lstlisting}[language=SQL, basicstyle=\small\ttfamily]
SELECT percentile(latency, 99) FROM requests
WHERE time > date_sub(curdate(), 2 WEEK)
AND app_version = "v8.2"
\end{lstlisting}
To process this query in time $t_\mathrm{query}$, we would need to merge $n_\mathrm{merge}$ summaries, each with runtime overhead $t_\mathrm{merge}$, and then estimate the quantile from the merged summary with runtime cost $t_\mathrm{est}$. This results in total query time:
\begin{equation}
t_\mathrm{query} = t_\mathrm{merge} \cdot n_\mathrm{merge} + t_\mathrm{est}.
\end{equation}
We evaluate the different regimes where queries are bottlenecked on merges and estimation in Figure~\ref{fig:varymerge} in Section~\ref{sec:evalmergeest}:
merge time begins to dominate at around $n_\mathrm{merge} \geq 10^4$.

We also consider \emph{threshold} queries which are conditioned on sub-groups or windows with percentiles above a specified threshold.
For example, we may be interested in combinations of application version and hardware platform for which the 99th percentile latency exceeds 100ms:
\begin{lstlisting}[language=SQL, basicstyle=\small\ttfamily]
SELECT app_version, hw_model,
    PERCENTILE(latency, 99) as p99 
FROM requests 
GROUP BY app_version, hw_model
HAVING p99 > 100
\end{lstlisting}
Such queries are very useful for debugging and data exploration~\cite{bailis2017macrobase}, but have additional runtime cost that depends on the number of groups $n_\mathrm{groups}$ since $t_\mathrm{est}$ can be significant when one is searching for high quantiles over thousands of sub-groups. This results in total query time:
\begin{equation}
t_\mathrm{query} = t_\mathrm{merge} \cdot n_\mathrm{merge} + t_\mathrm{est} \cdot n_\mathrm{groups}.
\end{equation}
\section{The Moments Sketch}
\label{sec:algorithm}
In this section, we describe how we perform quantile estimation using the \msketch.
First, we review the summary statistics stored in the \msketch and describe how they comprise an efficiently mergeable sketch. 
Second, we describe how we can use the method of moments and the maximum entropy principle to estimate quantiles from the \msketch, with details on how we resolve practical difficulties with numerical stability and estimation time.
We conclude with a discussion of theoretical guarantees on the approximation error of quantiles estimated from the sketch.

\subsection{Moments Sketch Statistics}
\label{sec:moments}
\begin{figure}
\includegraphics[width=\columnwidth]{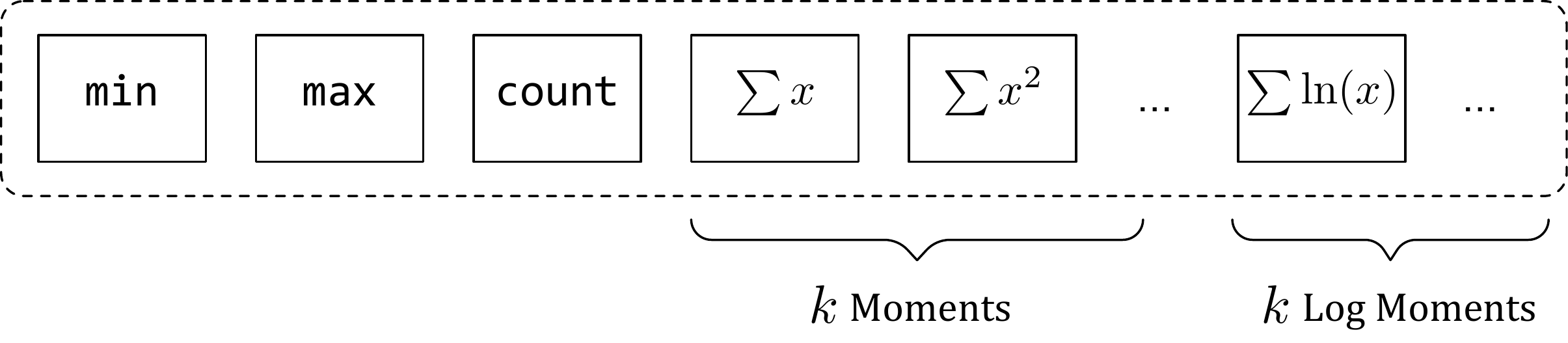}
\vspace{-2em}
\caption{The \msketch is an array of floating point values.}
\label{fig:datastructure}
\end{figure}

The \msketch of a dataset $D$ is a set of floating point values: the minimum value $x_\mathrm{min}$, the maximum value $x_\mathrm{max}$, the count $n$, the sample moments 
$\mu_i = \frac{1}{n}\sum_{x \in D} x^i$ \ and the sample logarithmic moments
$\nu_i = \frac{1}{n}\sum_{x \in D} \log^i{\left(x\right)}$ for $i \in \{1, \dots, k\}$ 
(Figure~\ref{fig:datastructure}).
The \msketch has an integer parameter $k$, which describes the highest power used in the moments.
We refer to $k$ as the \emph{order} of a \msketch.
Each sample moment provides additional information about the distribution, so higher-order \msketches are more precise but have higher space and computation time overheads.

\begin{algorithm}[t]
  \DontPrintSemicolon
  \SetAlgoLined
  \SetAlgoNoEnd
  
  \newcommand\mycommfont[1]{\rmfamily{#1}}
  \SetCommentSty{mycommfont}
  \SetKwComment{Comment}{$\triangleright$ }{}
  
  \Input{number of moments~$k$}
  
  \SetKwFunction{FInit}{Init}
  \Fn{\FInit{$x$}}{
      $x_\mathrm{min}, x_\mathrm{max}\leftarrow \infty, -\infty$\;
      $\vec{\mu}, \vec{\nu}, n \leftarrow \vec{0}, \vec{0}, n$ \;
  }
  
  \SetKwFunction{FUpdate}{Accumulate}
  \Fn{\FUpdate{$x$}}{
    $x_\mathrm{min},x_\mathrm{max} \leftarrow \min\{ x, x_\mathrm{min} \} , \max\{ x, x_\mathrm{max} \}$\;
    $n \leftarrow n + 1 $\;
    \For{$i \in \{1, \dots, k\}$}{
      $\mu_i \leftarrow \frac{n-1}{n}\mu_i + \frac{1}{n}x^i$ \Comment*[f]{Standard moments}
      
      \If{$x > 0$}{
        $\nu_i \leftarrow \frac{n-1}{n}\nu_i + \frac{1}{n}\log^i(x)$ \Comment*[f]{Log-moments}
      }
    }
  }

  \SetKwFunction{FMerge}{Merge}
  \Fn{\FMerge{$o$}\Comment*[f]{$o$ another sketch}}
  {
    $x_\mathrm{min} \leftarrow \min\{ o.x_\mathrm{min}, x_\mathrm{min} \} $\;
    $x_\mathrm{max} \leftarrow \max\{ o.x_\mathrm{max}, x_\mathrm{max} \}$ \;
    $\vec{\mu},\vec{\nu},n \leftarrow \vec{\mu}+o.\vec{\mu}, \vec{\nu}+o.\vec{\nu}, n+o.n$
  }
    
  \caption{Moments sketch operations}
  \label{alg:update}
\end{algorithm}

The \msketch supports a number of basic operations: \texttt{init} creates an empty sketch, \texttt{accumulate} updates the sketch via point-wise additions, and \texttt{merge} updates the sketch by merging it with another \msketch.
One can construct a \msketch over a dataset using either \texttt{accumulate} or \texttt{merge}.
When accumulating elements point-wise, we update the minimum and maximum, then add to the counts and moments.
As an implementation detail, we can accumulate the unscaled sums $\sum x^i$ and $\sum \log^i(x)$ instead of the $\mu_i,\nu_i$.
We merge two \msketches by combining the minimum, maximum, count, and the moments via comparison and potentially vectorized addition.
This merge operation preserves the property that a \msketch constructed using only \texttt{accumulate} is identical (up to floating point precision) to a \msketch constructed from merging existing sketches of partitions of the data, so there is no accuracy loss in pre-aggregating.
We provide pseudocode for these in Algorithm~\ref{alg:update}.
The \msketch additionally supports quantile estimation routines described in Section~\ref{sec:maxent} in order to answer end-user queries.
The \msketch thus supports all of the basic user-defined aggregate operations~\cite{cohen2006user,rusu2012glade} and can be incorporated into data systems using this API.

\minihead{Log-moments} The \msketch records logarithmic moments (log-moments) in order to recover better quantile estimates for long-tailed datasets.
In particular, taking the logarithm of data points is useful when values in the dataset can vary over several orders of magnitude.
In general, when updating a \msketch in a streaming manner or when maintaining multiple \msketches in a distributed setting, we cannot know \emph{a priori} whether standard moments or log-moments are more appropriate for the given dataset.
Therefore, our default approach is to store both sets of moments up to the same order $k$. Given additional prior knowledge of the data distribution, we may choose to maintain a \msketch with only a single set of moments.

Data points with negative values pose a potential problem for the log-moments since the logarithm is undefined for these points. 
There are several strategies for addressing this, including storing separate sums for positive and negative values and shifting all values so that they are positive.
In this paper, we adopt the simple approach of ignoring the log sums when there are any negative values, and computing estimates using the remaining statistics.

\minihead{Remark on pathological distributions}
The moments of certain ``pathological'' distributions may be undefined; for example, the Cauchy distribution $f(x) = \pi^{-1} \left(1 + x^2\right)^{-1}$ does not have finite moments of any order. 
However, the \msketch tracks the moments of an empirical dataset, which are always well-defined.
This suits our goal of estimating quantiles of a finite dataset, rather than an underlying distribution.

\subsection{Estimating Quantiles}
\label{sec:maxent}

\minihead{Method of moments} To estimate quantiles from a \msketch, we apply the \emph{method of moments}~\cite{wasserman2010all,anandkumar2012method,kalai2010efficiently,belkin2010polynomial} to construct a distribution $f(x)$ whose moments match those recorded in the sketch.
Specifically, given a \msketch with minimum $x_\mathrm{min}$ and maximum $x_\mathrm{max}$, we solve for a pdf $f(x)$ supported on $[x_\mathrm{min}, x_\mathrm{max}]$ with moments equal to the values in the \msketch. 
$$\int_{x_\mathrm{min}}^{x_\mathrm{max}} x^i f(x)\,dx = \mu_i \qquad \int_{x_\mathrm{min}}^{x_\mathrm{max}} \log^i(x) f(x) \, dx = \nu_i$$
We can then report the quantiles of $f(x)$ as estimates for the quantiles of the dataset.

In general, a finite set of moments does not uniquely determine a distribution~\cite{akhiezer1965classical}.
That is, there are often many possible distributions with varying quantiles that each match a given set of sample moments.
Therefore, we must disambiguate between them.

\minihead{Maximum entropy} In this work, we use the \emph{principle of maximum entropy}~\cite{jaynes1957} to select a unique distribution that satisfies the given moment constraints.
Intuitively, the differential Shannon entropy $H$ of a distribution with pdf $f(x)$, defined as
$H[f] = -\int_\mathcal{X} f(x)\log{f(x)}\,dx$,
is a measure of the degree of \emph{uninformativeness} of the distribution.
For example, a uniform distribution has higher entropy than a point mass distribution.
Thus, the maximum entropy distribution can be seen as the distribution that encodes the \emph{least} additional information about the data beyond that captured by the given moment constraints. 

Applying the maximum entropy principle to the \msketch, we estimate quantiles by solving for the pdf $f$ that maximizes the entropy while matching the moments in the sketch.
Following, we estimate quantiles using numeric integration and the Brent's method for root finding \cite{press2007numericalrecipes} .

In practice, we find that the use of maximum entropy distributions yields quantile estimates with comparable accuracy to alternative methods on a range of real-world datasets, unless the datasets are more discrete than continuous. We discuss our empirical results further in Section~\ref{sec:evalacc}.

\minihead{Optimization} We now describe how to solve for the maximum entropy distribution $f$.
We trade off between accuracy and estimation time by solving for $f$ subject to the first $k_1$ standard moments and $k_2$ log-moments stored in the sketch; incorporating more moments leads to more precise estimates but more computationally expensive estimation.
As previously noted, for datasets with non-positive values (i.e., $x_\mathrm{min} \leq 0$), we set $k_2 = 0$.
Therefore, our goal is to find the solution $f$ to the following optimization problem:
\begin{align}
  \underset{f\in\mathcal{F}[x_\mathrm{min}, x_\mathrm{max}]}{\mathrm{maximize}}& H[f] \label{eq:maxent-problem}\\
  \text{subject to } & \int_{x_\mathrm{min}}^{x_\mathrm{max}} x^i f(x)\,dx = \mu_i, & i \in \{1, \dots, k_1\} \nonumber \\
  	& \int_{x_\mathrm{min}}^{x_\mathrm{max}} \log^i(x) f(x) \, dx = \nu_i, & i \in \{1, \dots, k_2\} \nonumber
\end{align}
where $\mathcal{F}[x_\mathrm{min}, x_\mathrm{max}]$ denotes the set of distributions supported on $[x_\mathrm{min}, x_\mathrm{max}]$.

It is well known that the solution to Problem~\eqref{eq:maxent-problem} is a member of the class of exponential family distributions \cite{jaynes1957}:
\begin{equation*}
  f(x;\theta) = \exp\left(\theta_0 + \sum_{i=1}^{k_1} \theta_i x^i + \sum_{i=1}^{k_2} \theta_{k_1 + i} \log^i(x) \right),
\end{equation*}
where $\theta_0$ is a normalization constant such that $f(x;\theta)$ integrates to $1$ over the domain $[x_\mathrm{min}, x_\mathrm{max}]$. 
The maximum entropy distribution is determined by the parameter $\theta$ such that $f(x;\theta)$ satisfies the moment constraints in Problem~\eqref{eq:maxent-problem}. 

In order to solve for $\theta$, we define the potential function $L(\theta)$ from~\cite{mead1984maxent}:
\begin{align}
L(\theta) =& \int_{x_\mathrm{min}}^{x_\mathrm{min}} \exp{\left(\sum_{i=0}^{k_1} \theta_i x^i + \sum_{i=1}^{k_2} \theta_{k_1+i} \log^i{x} \right)}dx \label{eqn:potential} \\
 & \quad - \theta_0 - \sum_{i=0}^{k_1} \theta_i \mu_i -\sum_{i=1}^{k_2} \theta_{k_1+i} \nu_i \nonumber 
\end{align}
$L(\theta)$ is a convex function over $\theta$ and is constructed so that the minimizing solution $\theta_\text{opt} = \arg\min_{\theta \in \mathbb{R}^{k_1+k_2-1}}L(\theta)$ is exactly the set of coefficients which satisfy the constraints in Problem~\eqref{eq:maxent-problem}.
Equation~\eqref{eqn:potential} thus transforms the constrained optimization problem in~\eqref{eq:maxent-problem} into an unconstrained convex optimization problem which we solve using Newton's method~\cite{boydcvx}.
We show the explicit formulas for the gradient and Hessian that of Equation~\eqref{eqn:potential} in Appendix~\ref{sec:appendix_maxent} in~\cite{gan2018momentext}.
First-order optimization routines such as SGD and BFGS~\cite{Liu1989LBFGS} are also viable: they do not use the Hessian but require more steps to achieve convergence.
As we will describe in Section~\ref{sec:optimizations}, each additional entry in our Hessian can be computed efficiently using Chebyshev approximations, making second order methods more efficient overall.
We provide a lesion study comparison in Section~\ref{sec:est_lesion}.

\subsection{Practical Implementation}
\label{sec:optimizations}
In this section, we outline implementation concerns that are important for querying the \msketch in practice.
We include a number of optimizations to improve the stability and performance of Newton's method, and also discuss the stability of the \msketch under floating point precision.
Due to space constraints, some equations are omitted and provided in Appendix~\ref{sec:appendix_maxent} and~\ref{sec:appendix_stability} in~\cite{gan2018momentext}.

\subsubsection{Optimizing Newton's Method}
The primary source of difficulties in implementing Newton's method is the Hessian $\nabla^2 L$ of our objective $L$. In our case:
\begin{equation}
  \nabla^2 L(\theta)_{ij} = \int_{x_\mathrm{min}}^{x_\mathrm{max}} m_i(x) m_j(x) f(x; \theta) \, dx,
\label{eqn:hessian}
\end{equation}
where the functions $m_i(x)$ range over the set of functions
\begin{equation*}\{x^i : i \in \{1, \dots, k_1 \}\}\cup\{\log^i(x) : i \in \{1, \dots, k_2 \} \}.\end{equation*}
There are two main challenges in performing a Newton step using this Hessian.
First, $\nabla^2 L$ can be nearly singular and cause numerical instabilities in Newton's method that prevent or slow down convergence.
Second, since the integral in Eq.~\eqref{eqn:hessian} has no closed form, the cost of performing $O(k^2)$ numerical integrations to compute $\nabla^2 L$ in each iteration can be expensive. 

\minihead{Conditioning} To quantify the degree of numerical instability, we use the \emph{condition number} of the Hessian $\nabla^2 L$.
The condition number $\kappa(A)$ of a matrix $A$ describes how close a matrix is to being singular: matrices with high condition number are close to being singular, and $\log_{10} \kappa$ provides an estimate of how many digits of precision are lost when inverting $A$.
In particular, the use of the powers $m_i(x)\in\{x^i : i \in \{1, \dots, k_1 \}\}$ can result in ill-conditioned Hessians~\cite{Gautschi1978Condition}.
For example, when solving for a maximum entropy distribution with $k_1=8, k_2=0, x_\mathrm{min}=20$, and $x_\mathrm{max}=100$, we encountered $\kappa(\nabla^2 L) \approx 3 \cdot 10^{31}$ at $\theta=0$, making even the very first Newton step unstable.

We mitigate this issue by using a change of basis from the functions $m_i(x)=x^j$ and $m_i(x)=\log^j(x)$ to the basis of Chebyshev polynomials $T_i(x)$.
Chebyshev polynomials are bounded polynomials supported on $[-1,1]$ and are often used for polynomial approximation~\cite{badyo2005max,press2007numericalrecipes}.
Using them we define the new basis $\tilde{m}_i$ as follows:
\begin{equation*}
    \tilde{m}_i(x) = 
    \begin{cases}
        T_i(s_1(x)), &i \in\{1, \dots, k_1\} \\
        T_{i-k_1}(s_2(\log(x))), &i \in\{k_1 + 1, \dots, k_1 + k_2\}
    \end{cases}
\end{equation*}
where $s_1,s_2$ are linear scaling functions that map to $[-1,1]$.
The new basis functions $\tilde{m}_i(x)$ can be expressed in terms of $x^j$ and $\log^j(x)$ using standard formulae for Chebyshev polynomials and the binomial expansion \cite{mason2002chebyshev}.
Using this new basis for $m_i$ Equation~\eqref{eqn:hessian}, we found that the condition number for the above example is reduced to $\kappa \approx 11.3$, making precision loss during each Newton step less of a concern.

On certain datasets, if ill-conditioned matrices are still an issue at query time we further limit ourselves to using the first $k_1\leq k$ moments and $k_2\leq k$ log moments by selecting $k_1,k_2$ such that the condition number of the Hessian is less than a threshold $\kappa_\mathrm{max}$.
Our heuristics select $k_1,k_2$ by greedily incrementing $k_1$ and $k_2$ and favoring moments which are closer to the moments expected from a uniform distribution.

\minihead{Efficient Integration} Na\"{i}vely computing the Hessian in Equation~\eqref{eqn:hessian} requires evaluating $O(k^2)$ numerical integrals per iteration, which can lead to prohibitively slow estimation time.
We reduce this computational overhead by using polynomial approximations of the functions appearing in the integrands.
If the integrands $\tilde{m}_i(x)\tilde{m}_j(x) f(x;\theta)$ were expressible as polynomials in $x$, then each integral can be evaluated in closed form.
The factors in the integrand that do not appear as polynomials in $x$ are $\tilde{m}_i(x), i \in \{k_1 +1, \dots, k_1 + k_2 \}$, which are polynomials in $\log(x)$, and the pdf $f(x; \theta)$.
Therefore, we compute Chebyshev polynomial approximations of these factors and replace each instance in the integrands with its corresponding approximation.\footnote{Compare with Clenshaw-Curtis integration \cite{press2007numericalrecipes}.}

Approximating each of the factors with a degree $n_c$ polynomial takes $O(n_c \cdot \log{n_c})$ using a fast cosine transform~\cite{press2007numericalrecipes}, so computing the Hessian can be done in $O(k_2 n_c\log{n_c} + n_c k_1 k_2)$.
This is not an asymptotic improvement over naive numeric integration, but the number of complex function evaluations (i.e. $\cos(x), e^x$) is reduced substantially.
As we show in our empirical evaluation (Section~\ref{sec:est_lesion}), polynomial approximations reduce solve times $20\times$ compared to numerically integrating each entry of the Hessian independently.
We find in our experiments that the major bottleneck during maximum entropy optimization is the cosine transform used to construct the polynomial approximations.

\subsubsection{Floating point stability}
\label{sec:choosek}
Numeric floating point stability limits the range of useful $k$ in a \msketch.
Both our estimation routine and error bounds (Section~\ref{sec:errorgeneral}) use moments corresponding to data shifted onto the range $[-1,1]$.
On scaled data with range $[c-1,c+1]$, this leads to numeric error $\epsilon_k$ in the $k$-th shifted moment, bounded by $\epsilon_k \leq 2^k\left(|c|+1\right)^k\epsilon_s$ where $\epsilon_s$ is the relative error in the raw \msketch power sums.
This shift is the primary source of precision loss. 
We relate the loss to the error bounds in Section~\ref{sec:errorgeneral} to show that when using double precision floating point moments up to around $k \leq \frac{13.06}{0.78 + \log_{10}(|c|+1)}$ provide numerically useful values.
Data centered at 0 ($c=0$) have stable higher moments up to $k=16$, and in practice we encounter issues when $k \geq 16$.
We provide derivations and evaluations of this formula in Appendix~\ref{sec:appendix_stability} and~\ref{sec:eval_lowprecision} in~\cite{gan2018momentext}


\subsection{Quantile Error Bounds}
\label{sec:errorgeneral}
Recall that we estimate quantiles by constructing a maximum entropy distribution subject to the constraints recorded in a \msketch.
Since the true empirical distribution is in general not equal to the estimated maximum entropy distribution, to what extent can the quantiles estimated from the sketch deviate from the true quantiles?
In this section, we discuss worst-case bounds on the discrepancy between \emph{any} two distributions which share the same moments, and relate these to bounds on the quantile estimate errors.
In practice, error on non-adversarial datasets is lower than these bounds suggest.

We consider distributions supported on $[-1, 1]$: we can scale and shift any distribution with bounded support to match. 
By Proposition 1 in Kong~and~Valiant~\cite{kongvaliant2017spectrum}, any two distributions supported on $[-1, 1]$ with densities $f$ and $g$ and standard moments $\mu_f,\mu_g$, the Wasserstein distance (or Earth Mover's distance) $W_1(f, g)$ between $f$ and $g$ is bounded by:
\begin{equation*}
  W_1(f, g) \leq O\left(\frac{1}{k} + 3^k \|\mu_f - \mu_g\|_2 \right).
\end{equation*}

For univariate distributions $f$ and $g$, the Wasserstein distance between the distributions is equal to the L1 distance between their respective cumulative distribution functions $F$ and $G$ (see Theorem 6.0.2 in \cite{ambrosio2008gradient}). Thus:
\begin{equation*}
W_1(f,g) = \int_{-1}^{+1} |F(x) - G(x)| \,dx.
\label{eqn:cdfbound}
\end{equation*}

If $f$ is the true dataset distribution, we estimate $\hat{q}_\phi$ by calculating the $\phi$-quantile of the maximum entropy distribution $\hat{f}$.
The quantile error $\varepsilon(\hat{q}_\phi)$ is then equal to the gap between the CDFs: $\varepsilon(q_\phi) = |F(\hat{q}_\phi) - \hat{F}(\hat{q}_\phi)|$. 
Therefore, the average quantile error over the support $[-1, 1]$ is bounded as follows:
\begin{equation}
  \int_{-1}^{+1} \varepsilon(x)\,dx \leq O\left(\frac{1}{k} + 3^k \|\mu_f - \mu_{\hat{f}}\|_2\right).
  \label{eqn:avgerrorrange}
\end{equation}
Since we can run Newton's method until the moments $\mu_f$ and $\mu_{\hat{f}}$ match to any desired precision, the $3^k \|\mu_f - \mu_{\hat{f}}\|_2$ term is negligible.

Equation~\eqref{eqn:avgerrorrange} does not directly apply to the $\epsilon_\mathrm{avg}$ used in Section~\ref{sec:eval}, which is averaged over $\phi$ for uniformly spaced $\phi$-quantiles rather than over the support of the distribution.
Since $\phi=\hat{F}(\hat{q}_\phi)$,
we can relate $\epsilon_\mathrm{avg}$ to Eq.~\eqref{eqn:avgerrorrange} using our maximum entropy distribution $\hat{f}$:
$$
\epsilon_{\mathrm{avg}} = \int_{0}^{1} \varepsilon(\hat{q}_\phi)\,d\phi = \int_{-1}^{+1} \varepsilon(x)\hat{f}(x)\,dx \leq O\left(\frac{\hat{f}_\mathrm{max}}{k}\right)
$$
where $\hat{f}_\mathrm{max}$ is the maximum density of our estimate.
Thus, we expect the average quantile error $\epsilon_\mathrm{avg}$ to have a decreasing upper bound as $k$ increases, with higher potential error when $\hat{f}$ has regions of high density relative to its support.
Though these bounds are too conservative to be useful in practice, they provide useful intuition on how worst case error can vary with $k$ and $\hat{f}$ (Figure~\ref{fig:size_bound}).
\section{Threshold Queries}
\label{sec:alg_tresholds}
We described in Section~\ref{sec:costmodel} two types of queries: single quantile queries and threshold queries over multiple groups.
The optimizations in Section~\ref{sec:optimizations} can bring quantile estimation overhead down to $\leq 1$ms, which is sufficient for interactive latencies on single quantile queries.
In this section we show how we can further reduce quantile estimation overheads on threshold queries.
Instead of computing the quantile on each sub-group directly, we compute a sequence of progressively more precise bounds in a \emph{cascade}~\cite{viola2001rapid}, and only use more expensive estimators when necessary.
We first describe a series of bounds relevant to the \msketch in Section~\ref{sec:querybounds} and then show how they can be used in end-to-end queries in Section~\ref{sec:cascades}.

\subsection{Moment-based inequalities}
\label{sec:querybounds}
Given the statistics in a \msketch, we apply a variety of classical inequalities to derive bounds on the quantiles.
These provide worst-case error guarantees for quantile estimates, and enable faster query processing for threshold queries over multiple groups.

One simple inequality we make use of is Markov's inequality.
Given a non-negative dataset $D$ with moments $\mu_i$ Markov's inequality tells us that for any value $t$,
$\text{rank}(t) \geq n\left(1-\frac{\mu_k}{t^k}\right)$
where the $\text{rank}$ is the number of elements in $D$ less than $t$.
We can apply Markov's inequality to moments of transformations of $D$ including $T^{+}(D) = \{x - x_\mathrm{min}: x \in D\}$, $T^{-}(D) = \{x_\mathrm{max} - x : x \in D\}$, and $T^{l}(D)=\{\log(x) : x \in D\}$ to bound $\text{rank}(t)$ and thus also the error $\epsilon$ for quantile estimates $t=\hat{q}_\phi$. We refer to this procedure as the \markov procedure.

The authors in~\cite{racz2006bounding} provide a procedure \colora{(Section 3, Figure 1 in \cite{racz2006bounding})} for computing tighter but more computationally expensive bounds on the CDF $F(t)$ of a distribution given its moments.
We refer to this procedure as the \racz procedure, and as with the \markov procedure, use it to bound the error of a quantile estimate $\hat{q}_\phi$.
The \racz procedure does not make use of the standard moments and log moments simultaneously, so we run \racz once on the standard moments and once on log moments and take the tighter of the bounds.

\subsection{Cascades for Threshold queries}
\label{sec:cascades}
\begin{algorithm}[t]
  \DontPrintSemicolon
  \SetAlgoLined
  \SetAlgoNoEnd
  
  \newcommand\mycommfont[1]{\rmfamily{#1}}
  \SetCommentSty{mycommfont}
  \SetKwComment{Comment}{$\triangleright$ }{}

  \SetKwFunction{FCheckBound}{CheckBound}
  \Macro{\FCheckBound{$r_\mathrm{lower}$, $r_\mathrm{upper}$, $r_t$}}{
    \If{$r_\mathrm{lower} > r_t$}{
      \textbf{return} true
    }
    \ElseIf{$r_\mathrm{upper} < r_t$}{
      \textbf{return} false
    }
  }
  
  \SetKwFunction{FThreshold}{Threshold}
  \SetKwFunction{FMarkov}{MarkovBound}
  \SetKwFunction{FRTT}{RTTBound}
  \SetKwFunction{FMaxEnt}{MaxEntQuantile}
  \Fn{\FThreshold{\textnormal{threshold} $t$, \textnormal{quantile} $\phi$}}{
    \If{$t > x_\mathrm{max}$}{\textbf{return} false} 

    \If{$t < x_\mathrm{min}$}{\textbf{return} true}

    $r_\mathrm{lower},\, r_\mathrm{upper} \leftarrow \FMarkov{t}$ \Comment*[f]{Markov Bound}
    
    \FCheckBound{$r_\mathrm{lower}$, $r_\mathrm{upper}$, $n\phi$}\;
    $r_\mathrm{lower},\, r_\mathrm{upper} \leftarrow \FRTT{t}$ \Comment*[f]{RTT Bound}
    
    \FCheckBound{$r_\mathrm{lower}$, $r_\mathrm{upper}$, $n\phi$}\;
    $q_\phi \leftarrow \FMaxEnt{$\phi$}$ \Comment*[f]{Maximum Entropy}
    
    \textbf{return} $q_\phi > t$
  }
\caption{Threshold Query Cascade}
\label{alg:cascade}
\end{algorithm}

Given a \msketch, Algorithm~\ref{alg:cascade} shows how we calculate \texttt{Threshold}($t$, $\phi$): whether the dataset has quantile estimate $\hat{q}_\phi$ above a fixed cutoff $t$.
We use this routine whenever we answer queries on groups with a predicate $\hat{q}_\phi > t$, allowing us to check whether a subgroup should be included in the results without computing $\hat{q}_\phi$ directly.
The threshold check routine first performs a simple filter on whether the threshold $t$ falls in the range $[x_\mathrm{min},x_\mathrm{max}]$.
\colora{Then, we can use the Markov inequalities \texttt{MarkovBound} to calculate lower and upper bounds on the rank of the threshold $\text{rank}(t)$ in the subpopulation. 
Similarly the \texttt{RTTBound} routine uses more sophisticated inequalities in~\cite{racz2006bounding} to obtain tighter bounds on the rank.
These bounds are used to determine if we can resolve the threshold predicate immediately.
If not, we solve for the maximum entropy distribution as described in Section~\ref{sec:maxent} (\texttt{MaxEntQuantile}) and calculate $\hat{q}_\phi$.
}

The Markov and \racz bounds are cheaper to compute than our maximum entropy estimate, making threshold predicates cheaper to evaluate than explicit quantile estimates.
The bounds apply to any distribution or dataset that matches the moments in a \msketch, so this routine has no false negatives and is consistent with calculating the maximum entropy quantile estimate up front.
\section{Evaluation}
\label{sec:eval}
In this section we evaluate the efficiency and accuracy of the \msketch in a series of microbenchmarks, 
and then show how the \msketch provides end-to-end performance improvements in the Druid and Macrobase data analytics engines~\cite{bailis2017macrobase,Yang2014druid}.
 
This evaluation demonstrates that:
\begin{enumerate}[itemsep=.1em,parsep=.25em,topsep=.25em]
    \item The \msketch supports $15$ to $50\times$ faster query times than comparably accurate summaries on quantile aggregations.
    \item The \msketch provides $\epsilon_\mathrm{avg} \leq 0.01$ estimates across a range of real-world datasets using less than 200 bytes of storage.
    \item Maximum entropy estimation is more accurate than alternative moment-based quantile estimates, and our solver improves estimation time by $200\times$ over naive solutions.
    \item Integrating the \msketch as a user-defined sketch provides $7\times$ faster quantile queries than the default quantile summary in Druid workloads.
    \item Cascades can provide $25\times$ higher query throughput compared to direct \msketch usage in Macrobase threshold queries.
\end{enumerate}

Throughout the evaluations, the \msketch is able to accelerate a variety of aggregation-heavy workloads with minimal space overhead.

\subsection{Experimental Setup}
\label{sec:evalsetup}

We implement the \msketch and its quantile estimation routines in Java\footnote{\url{https://github.com/stanford-futuredata/msketch}}.
This allows for direct comparisons with the open source quantile summaries \cite{datasketches,druidblog} and integration with the Java-based Druid~\cite{Yang2014druid} and MacroBase~\cite{bailis2017macrobase} systems.
In our experimental results, we use the abbreviation M-Sketch to refer to the \msketch.

We compare against a number of alternative quantile summaries:
a mergeable equi-width histogram (\hist) using power-of-two ranges \cite{rabkin2014agg}, 
\colora{the `GKArray' (\gk) variant of the Greenwald Khanna \cite{luo2016quantiles,greenwald2001space} sketch,} 
the AVL-tree T-Digest (\tdigest) \cite{tdigest} sketch, 
the streaming histogram (\stree) in \cite{ben2010streaming} as implemented in Druid,
the `Random' (\random) sketch from \cite{luo2016quantiles,wang2013quantiles},
reservoir sampling (\sampling) \cite{Vitter1985random},
and the low discrepancy mergeable sketch (\yahoo) from \cite{agarwal2012mergeable}, 
both implemented in the Yahoo! datasketches library \cite{datasketches}.
The \gk sketch is not usually considered mergeable since its size can grow upon merging \cite{agarwal2012mergeable}, this is especially dramatic in the production benchmarks in Appendix~\ref{sec:appendix_aria} in~\cite{gan2018momentext}.
We do not compare against fixed-universe quantile summaries such as the Q-Digest~\cite{shrivastava2004medians} or Count-Min sketch~\cite{Cormode2005countmin} since they would discretize continuous values.

Each quantile summary has a \emph{size parameter} controlling its memory usage, which we will vary in our experiments.
Our implementations and benchmarks use double precision floating point values.
During \msketch quantile estimation we run Newton's method until the moments match to within $\delta=10^{-9}$, and select $k_1,k_2$ using a maximum condition number $\kappa_\mathrm{max} = 10^4$.
We construct the \msketch to store both standard and log moments up to order $k$, but choose at query time which moments to make use of as described in Section~\ref{sec:choosek}.

We quantify the accuracy of a quantile estimate using the quantile error $\varepsilon$ as defined in Section~\ref{sec:quantilebg}.
Then, as in~\cite{luo2016quantiles,wang2013quantiles} we can compare the accuracies of summaries on a given dataset by computing their \emph{average error} $\epsilon_\mathrm{avg}$ over a set of uniformly spaced $\phi$-quantiles. 
In the evaluation that follows, we test on 21 equally spaced $\phi$ between $0.01$ and $0.99$.

We evaluate each summary via single-threaded experiments on a machine with an Intel Xeon E5-4657L 2.40GHz processor and 1TB of RAM, omitting the time to load data from disk.

\subsubsection{Datasets}
We make use of six real-valued datasets in our experiments, whose characteristics are summarized in Table~\ref{tab:data_params}.
The \emph{milan} dataset consists of Internet usage measurements from Nov. 2013 in the Telecom Italia Call Data Records~\cite{milan}.
The \emph{hepmass} dataset consists of the first feature in the UCI~\cite{uci} HEPMASS dataset. 
The \emph{occupancy} dataset consists of CO2 measurements from the UCI Occupancy Detection dataset. 
The \emph{retail} dataset consists of integer purchase quantities from the UCI Online Retail dataset. 
The \emph{power} dataset consists of Global Active Power measurements from the UCI Individual Household Electric Power Consumption dataset.
The \emph{exponential} dataset consists of synthetic values from an exponential distribution with $\lambda=1$.

\begin{table}[]
\iftoggle{arxiv}{\small}{\scriptsize}
\centering
\begin{tabular}{@{}lllllll@{}}
\toprule
       & milan     & hepmass  & occupancy & retail & power  & expon \\
\midrule
size  & 81M & 10.5M & 20k & 530k & 2M & 100M \\
min   & $2.3e{-6}$ & $-1.961$ & $412.8$   & $1$    & $0.076$& $1.2e{-7}$\\
max   & $7936$     & $4.378$  & $2077$    & $80995$& $11.12$& $16.30$\\
mean  & $36.77$    & $0.0163$ & $690.6$   & $10.66$& $1.092$& $1.000$\\
stddev& $103.5$    & $1.004$  & $311.2$   & $156.8$& $1.057$& $0.999$\\
skew  & $8.585$    & $0.2946$ & $1.654$   & $460.1$& $1.786$& $1.994$\\
\bottomrule
\end{tabular}
\caption{\colora{Dataset Characteristics. The evaluation datasets cover a range of distribution types.}}
\vspace{-2em}
\label{tab:data_params}
\end{table}

\subsection{Performance Benchmarks}
We begin with a series of microbenchmarks evaluating the \msketch query times and accuracy.

\subsubsection{Query Time}
\label{sec:evalquerytime}
Our primary metric for evaluating the \msketch is total query time.
We evaluate quantile aggregation query times by pre-aggregating our datasets into cells of 200 values and maintaining quantile summaries for each cell.
Then we measure the time it takes to performing a sequence of merges and estimate a quantile.
In this performance microbenchmark, the cells are grouped based on their sequence in the dataset, while the evaluations in Section~\ref{sec:evalapps} group based on column values.
\colora{We divide the datasets into a large number of cells to simulate production data cubes, while in Appendix~\ref{sec:appendix_cellsize} and~\ref{sec:appendix_aria} in~\cite{gan2018momentext} we vary the cell sizes.
Since the \msketch size and merge time are data-independent, the results generalize as we vary cell size.}

\begin{figure}
    \includegraphics[width=\columnwidth]{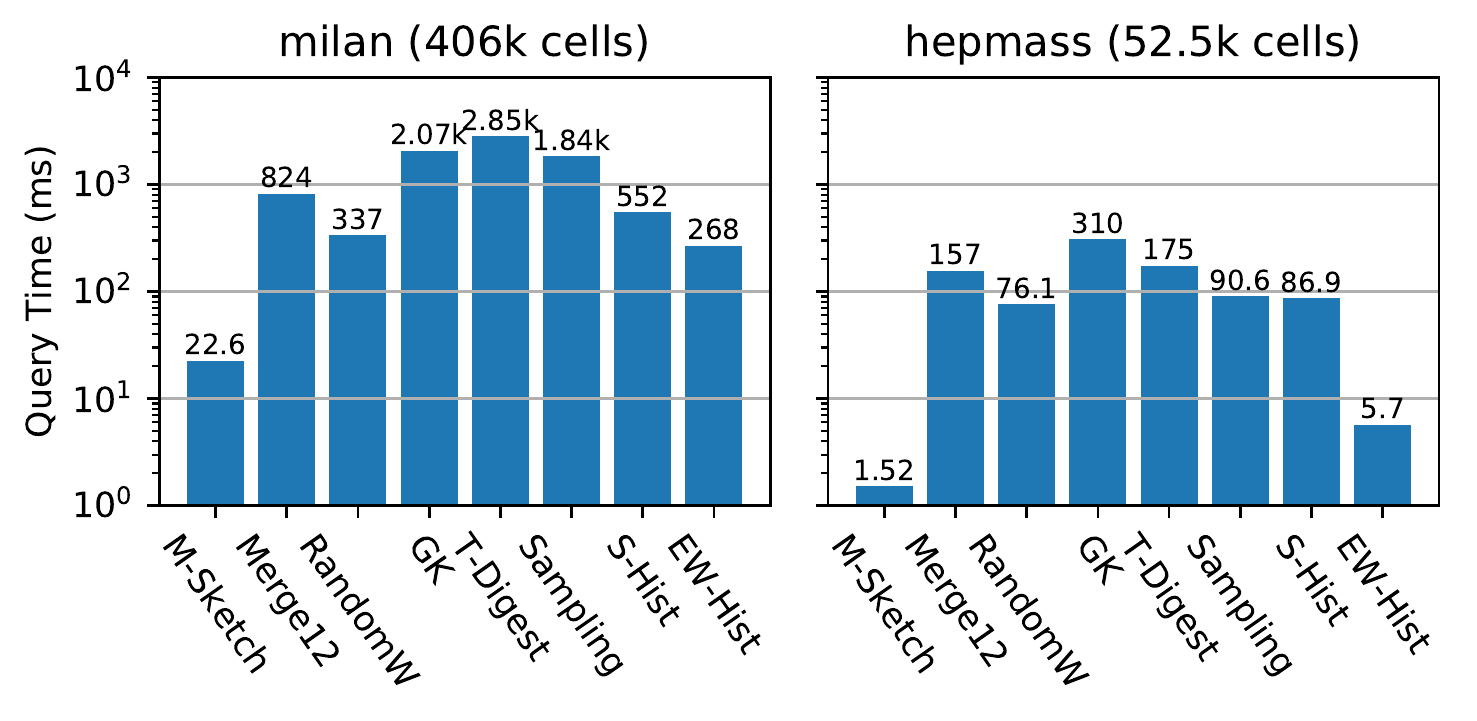}
    \vspace{-2em}
    \caption{Total query time using different summaries to estimate quantiles with $\epsilon_\mathrm{avg} \leq .01$. The \msketch enables significantly faster queries at this accuracy.}
    \label{fig:querytime}
\end{figure}
Figure~\ref{fig:querytime} shows the total query time to merge the summaries and then compute a quantile estimate 
when each summary is instantiated at the smallest size sufficient to achieve $\epsilon_\mathrm{avg} \leq .01$ accuracy.
We provide the parameters we used and average observed space usage in Table~\ref{tab:sketch_params}.
On the long-tailed milan dataset, the \stree and \hist summaries are unable to achieve $\epsilon_\mathrm{avg} \leq .01$ accuracy with less than 100 thousand buckets, so we provide timings at 100 buckets for comparison.
The \msketch provides $15$ to $50\times$ faster query times than \random, the next fastest accurate summary.
\colora{As a baseline, sorting the milan dataset takes 7.0 seconds, selecting an exact quantile online takes 880ms, and streaming the data pointwise into a \random sketch with $\epsilon=1/40$ takes 220ms.
These methods do not scale as well as using pre-aggregated \msketches as dataset density grows but the number of cells remains fixed.}

\begin{table}[]
\iftoggle{arxiv}{\small}{\scriptsize}
\centering
\begin{tabular}{@{}lllll@{}}
\toprule
dataset    & milan &                 & hepmass &     \\ 
sketch     & param & size (b) & param & size (b) \\
\midrule
\texttt{M-Sketch}  & $k=10$                  & 200  
& $k=3$   & 72     \\
\yahoo    & $k=32$                   & 5920  
& $k=32$  & 5150          \\
\random   & $\epsilon=\frac{1}{40}$ & 3200 & $\epsilon=\frac{1}{40}$ & 3375 \\
\gk       & $\epsilon=\frac{1}{60}$ & 720  
& $\epsilon=\frac{1}{40}$ & 496\\
\tdigest  & $\delta=5.0$                  & 769  
& $\delta=1.5$ &  93\\
\sampling & 1000 samples                  & 8010  
& 1000    & 8010  \\
\stree    & 100 bins                      & 1220  
& 100     & 1220 \\
\hist     & 100 bins                      & 812  
& 15      & 132  \\ \bottomrule
\end{tabular}
\caption{Summary size parameters used in Figure~\ref{fig:querytime}. We use these parameters to compare the query times at $\epsilon_\mathrm{avg} \leq .01$ accuracy.}
\vspace{-2em}
\label{tab:sketch_params}
\end{table}
    
\subsubsection{Merge and Estimation Time}
\label{sec:evalmergeest}
Recall that for a basic aggregate quantile query
$t_\mathrm{query} = t_\mathrm{merge} \cdot n_\mathrm{merge} + t_\mathrm{est}.$
Thus we also measure $t_\mathrm{merge}$ and $t_\mathrm{est}$ to quantify the regimes where the \msketch performs well.
In these experiments, we vary the summary size parameters, though many summaries have a minimum size, and the \msketch runs into numeric stability issues past $k \geq 15$ on some datasets (see Section~\ref{sec:choosek}).

\begin{figure}
    \includegraphics[width=\columnwidth]{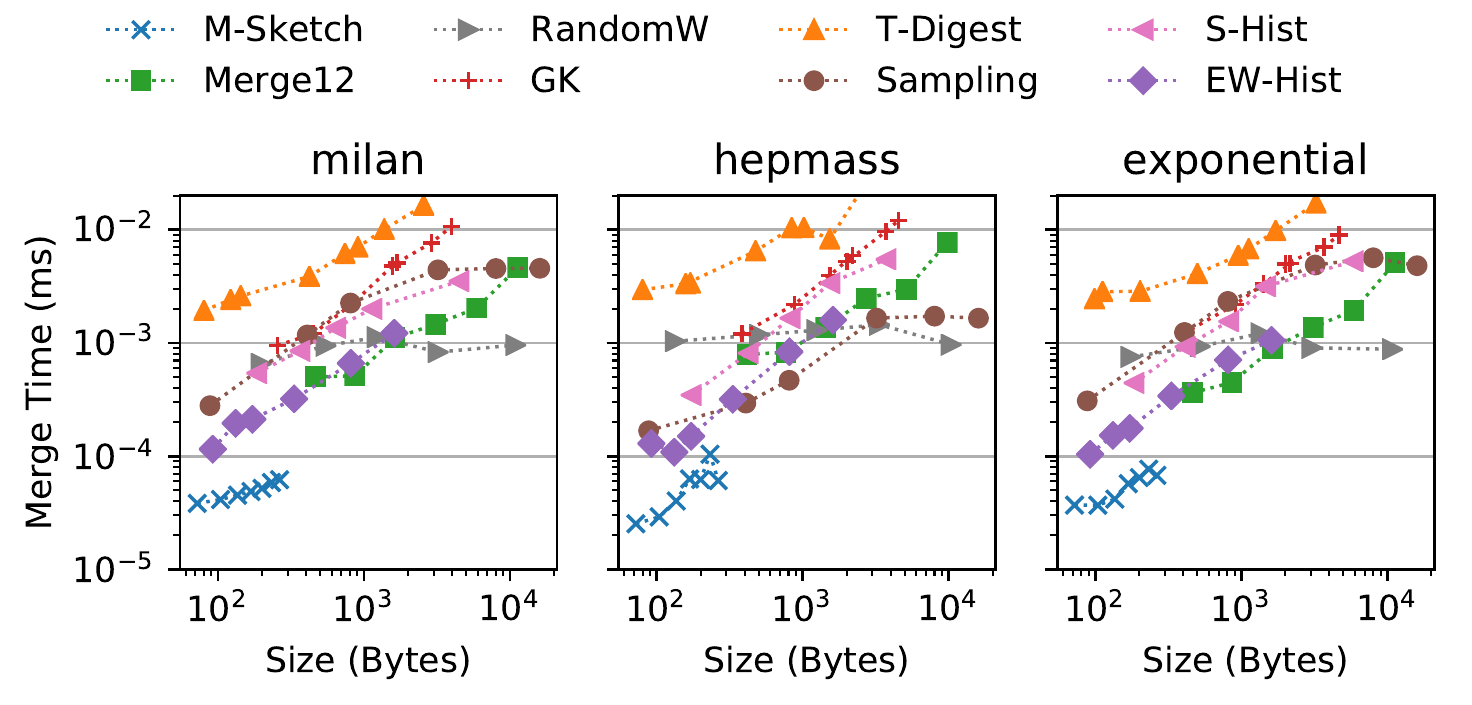}
    \vspace{-2em}
    \caption{Per-merge latencies. The \msketch provides faster merge times than alternative summaries at the same size.}
    \label{fig:mergetime}
    \vspace{-1em}
\end{figure}

In Figure~\ref{fig:mergetime} we evaluate the average time required to merge one of the cell summaries.
Larger summaries are more expensive to merge, and the \msketch has faster ($< 50$ns) merge times throughout its size range.
When comparing summaries using the parameters in Table~\ref{tab:sketch_params}, the \msketch has up to $50\times$ faster merge times than other summaries with the same accuracy.

\begin{figure}
    \includegraphics[width=\columnwidth]{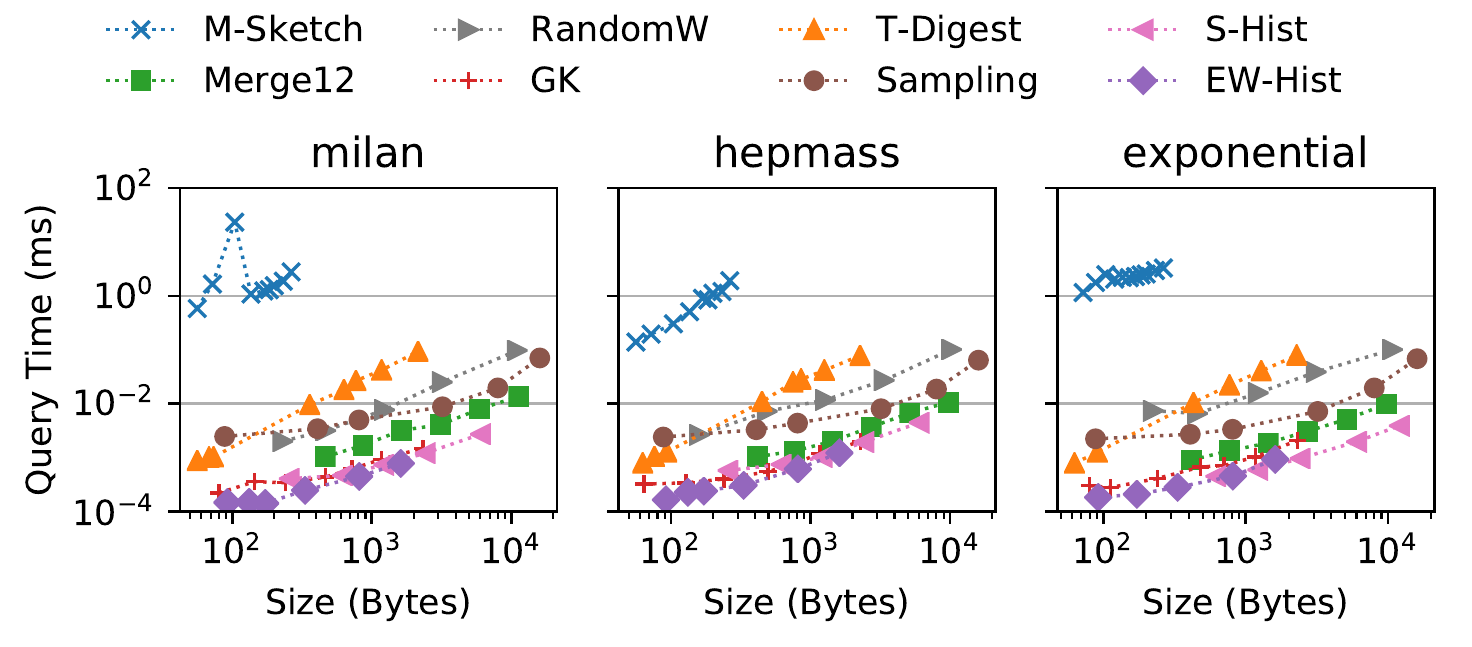}
    \vspace{-2em}
    \caption{Quantile Estimation time. Estimation time on the \msketch is slower than other sketches but under 3ms for $k=10$.}
    \vspace{-1em}
    \label{fig:estimationtime}
\end{figure}

\begin{figure}
    \includegraphics[width=\columnwidth]{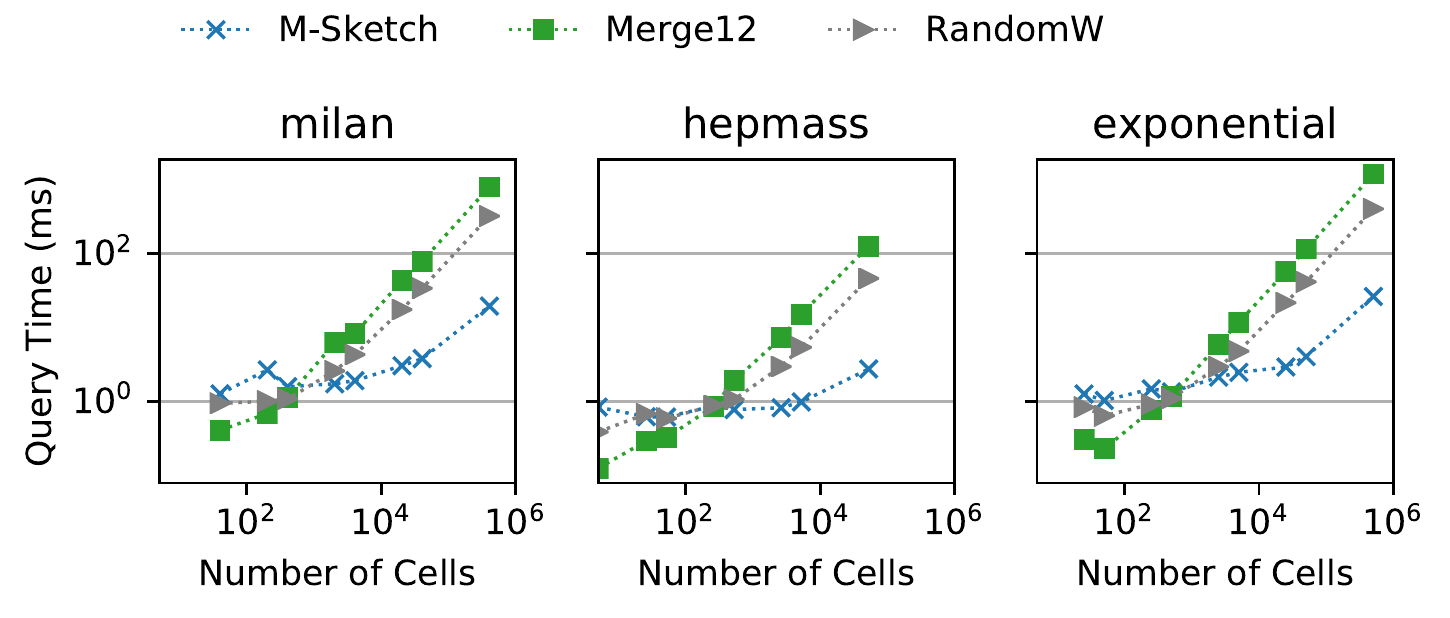}
    \vspace{-2em}
    \caption{Comparing total query time using different mergeable summaries as we vary the number of merges. The \msketch provides performance improvements when $n_\mathrm{merge} \geq 10^4$.}
    \label{fig:varymerge}
\end{figure}

One can also parallelize the merges by sharding the data and having separate nodes operate over each partition, generating partial summaries to be aggregated into a final result.
Since each parallel worker can operate independently, in these settings the \msketch maintains the same relative performance improvements over alternative summaries when we can amortize fixed overheads, and we include supplemental parallel experiments in Appendix~\ref{sec:appendix_parallel} in~\cite{gan2018momentext}.
The other major contributor to query time is estimation time. 
In Figure~\ref{fig:estimationtime} we measure the time to estimate quantiles given an existing summary. 
The \msketch provides on average 2 ms estimation times, though estimation time can be higher when our estimator chooses higher $k_1,k_2$ to achieve better accuracy.
This is the cause for the spike at $k=4$ in the milan dataset and users can can mitigate this by lowering the condition number threshold $\kappa_\mathrm{max}$.
Other summaries support microsecond estimation times.
The \msketch thus offers a tradeoff of better merge time for worse estimation time.
If users require faster estimation times, the cascades in Section~\ref{sec:cascades} and the alternative estimators in Section~\ref{sec:est_lesion} can assist.
We show how the merge time and estimation time tradeoff define regimes where each component dominates depending on the number of merges.
In Figure~\ref{fig:varymerge} we measure how the query time changes as we vary the number of summaries (cells of size 200) we aggregate.
We use the \msketch with $k=10$ and compare against the mergeable \yahoo and \random summaries with parameters from Table~\ref{tab:sketch_params}.
When $n_\mathrm{merge} \geq 10^4$ merge time dominates and the \msketch provides better performance than alternative summaries.
However, the \msketch estimation times dominate when $n_\mathrm{merge} \leq 100$.

\subsubsection{Accuracy}
\label{sec:evalacc}
The \msketch accuracy is dataset dependent, so in this section we compare the average quantile error on our evaluation datasets.

\begin{figure}
    \includegraphics[width=\columnwidth]{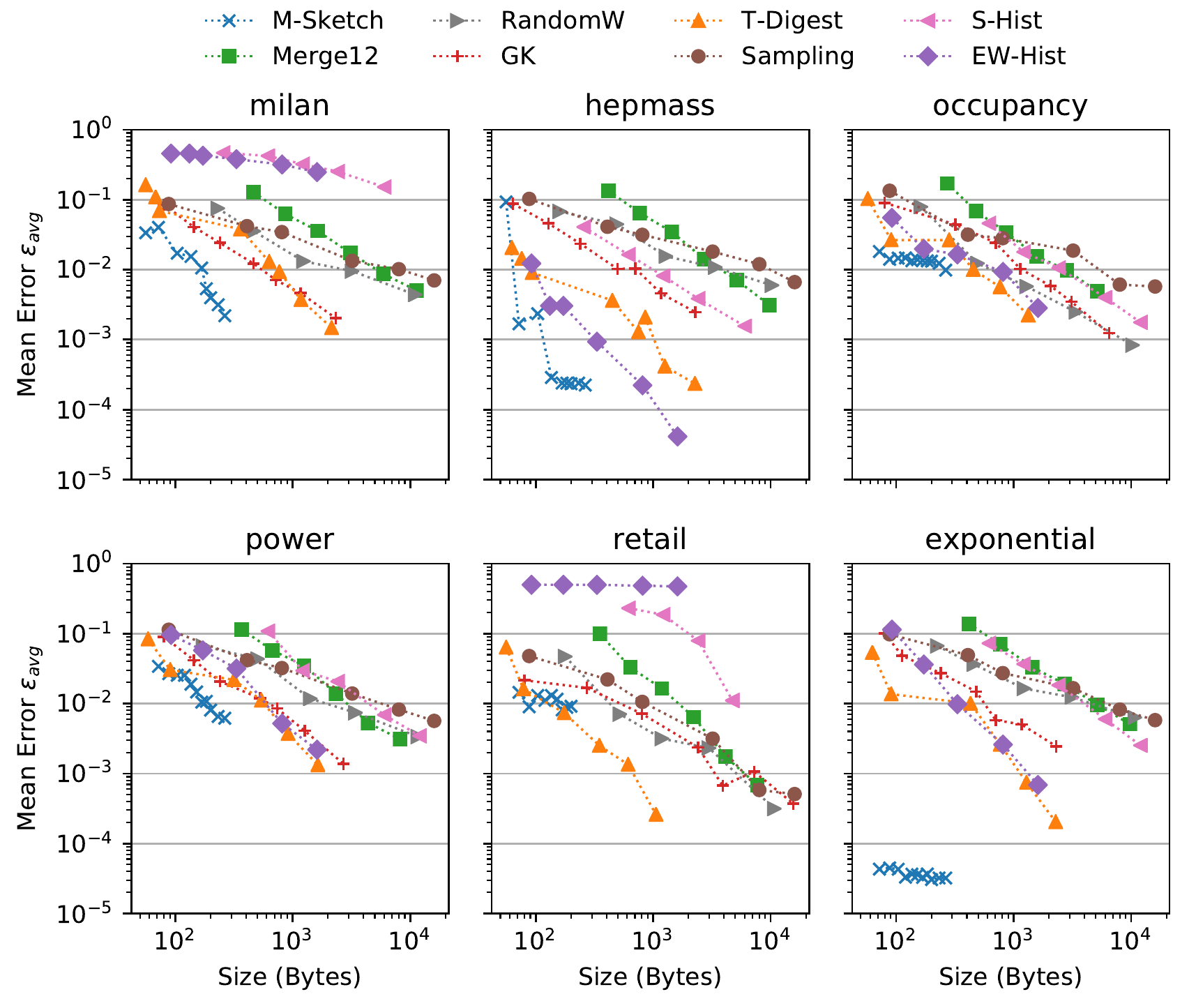}
    \vspace{-2em}
    \caption{Average error for summaries of different sizes. The \msketch delivers consistent $\epsilon_\mathrm{avg} \leq 0.015$ with fewer than 200 bytes.}
    \label{fig:size_accuracy}
\end{figure}

Figure~\ref{fig:size_accuracy} illustrates the average quantile error $\epsilon_\mathrm{avg}$ for summaries of different sizes constructed using pointwise accumulation on the complete dataset.
The \msketch achieves $\epsilon \leq 10^{-4}$ accuracy on the synthetic exponential dataset, and $\epsilon \leq 10^{-3}$ accuracy on the high entropy hepmass dataset. 
On other datasets it is able to achieve $\epsilon_\mathrm{avg} \leq 0.01$ with fewer than 200 bytes of space.
On the integer retail dataset we round estimates to the nearest integer.
The \hist summary, while efficient to merge, provides less accurate estimates than the \msketch, especially in the long-tailed milan and retail datasets.

\begin{figure}
    \includegraphics[width=\columnwidth]{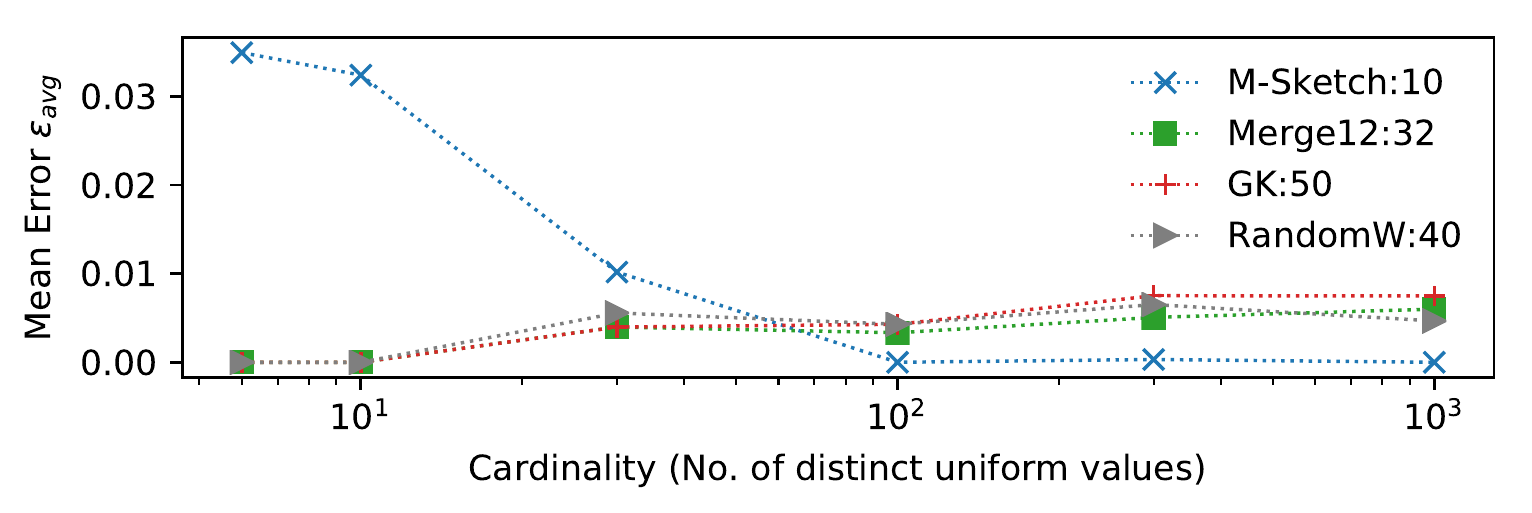}
    \vspace{-2em}
    \caption{\colora{Accuracy of maximum entropy estimates on distributions with varying cardinality. The \msketch is less accurate on discretized datasets, and fails to converge for cardinalities $n < 5$.}}
    \label{fig:robustdiscrete}
    \vspace{-1em}
\end{figure}

We provide further experiments in~\cite{gan2018momentext} showing how the \msketch worst-case error bounds are comparable to other summaries (Appendix~\ref{sec:errorboundeval}), that the \msketch is robust to changes in skew and the presence of outliers (Appendix~\ref{sec:varydistribution}), and that the \msketch generalizes to a production workload (Appendix~\ref{sec:appendix_aria}).
However, on datasets with low-entropy, in particular datasets consisting of a small number of discrete point masses, the maximum entropy principle provides poor accuracy.
In the worst case, the maximum entropy solver can fail to converge on datasets with too few distinct values.
Figure~\ref{fig:robustdiscrete} illustrates how the error of the maximum entropy estimate increases as we lower the cardinality of a dataset consisting of uniformly spaced points in the range $[-1,1]$, eventually failing to converge on datasets with fewer than five distinct values.
If users are expecting to run queries on primarily low-cardinality datasets, fixed-universe sketches or heavy-hitters sketches may be more appropriate.

\subsection{Quantile Estimation Lesion Study}
\label{sec:est_lesion}
To evaluate each component of our quantile estimator design, we compare the accuracy and estimation time of a variety of alternative techniques on the milan and hepmass datasets.
We evaluate the impact of using log moments, the maximum entropy distribution, and our optimizations to estimation.

\begin{figure}
    \includegraphics[width=\columnwidth]{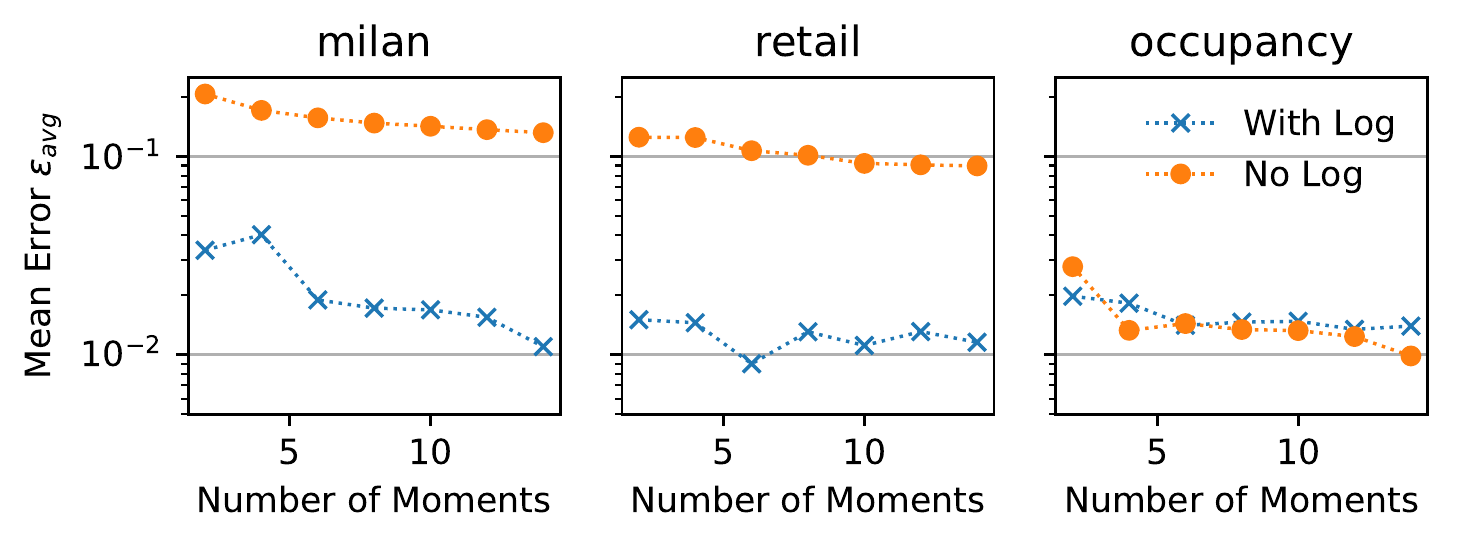}
    \vspace{-2em}
    \caption{Accuracy with and without log moments. Given the same total space budget, log moments improve accuracy on the long-tailed milan and retail datasets, and do not affect accuracy significantly on other datasets such as occupancy}
    \vspace{-1em}
    \label{fig:loglesion}
\end{figure}
To examine effectiveness of log moments, we compare our maximum entropy quantile estimator accuracy with and without log moments.
For a fair comparison, we compare the estimates produced from $k$ standard moments and no log moments with those produced from up to $\frac{k}{2}$ of each.
Figure~\ref{fig:loglesion} illustrates how on some long-tailed datasets, notably milan and retail, log moments reduce the error from $\epsilon > .15$ to $\epsilon < .015$.
On other datasets, log moments do not have a significant impact.

We compare our estimator (\texttt{opt}) with a number of other estimators that make use of the same moments.
The gaussian estimator fits a Gaussian distribution to the mean and standard deviation.
The mnat estimator uses the closed form discrete CDF estimator in~\cite{MNATSAKANOV2008hausdorff}.
The svd estimator discretizes the domain and uses singular value decomposition to solve for a distribution with matching moments.
The cvx-min estimator also discretizes the domain and uses a convex solver to construct a distribution with minimal maximum density and matching moments.
The cvx-maxent estimator discretizes the domain and uses a convex solver to maximize the entropy, as described in Chapter 7 in~\cite{boydcvx}.
The newton estimator implements our estimator without the integration techniques in Sec.~\ref{sec:optimizations}, and uses adaptive Romberg integration instead \cite{press2007numericalrecipes}.
\colora{The bfgs estimator implements maximum entropy optimization using the first-order L-BFGS~\cite{Liu1989LBFGS} method as implemented in a Java port of liblbfgs~\cite{lbfgs4j}.}

\begin{figure}
    \includegraphics[width=\columnwidth]{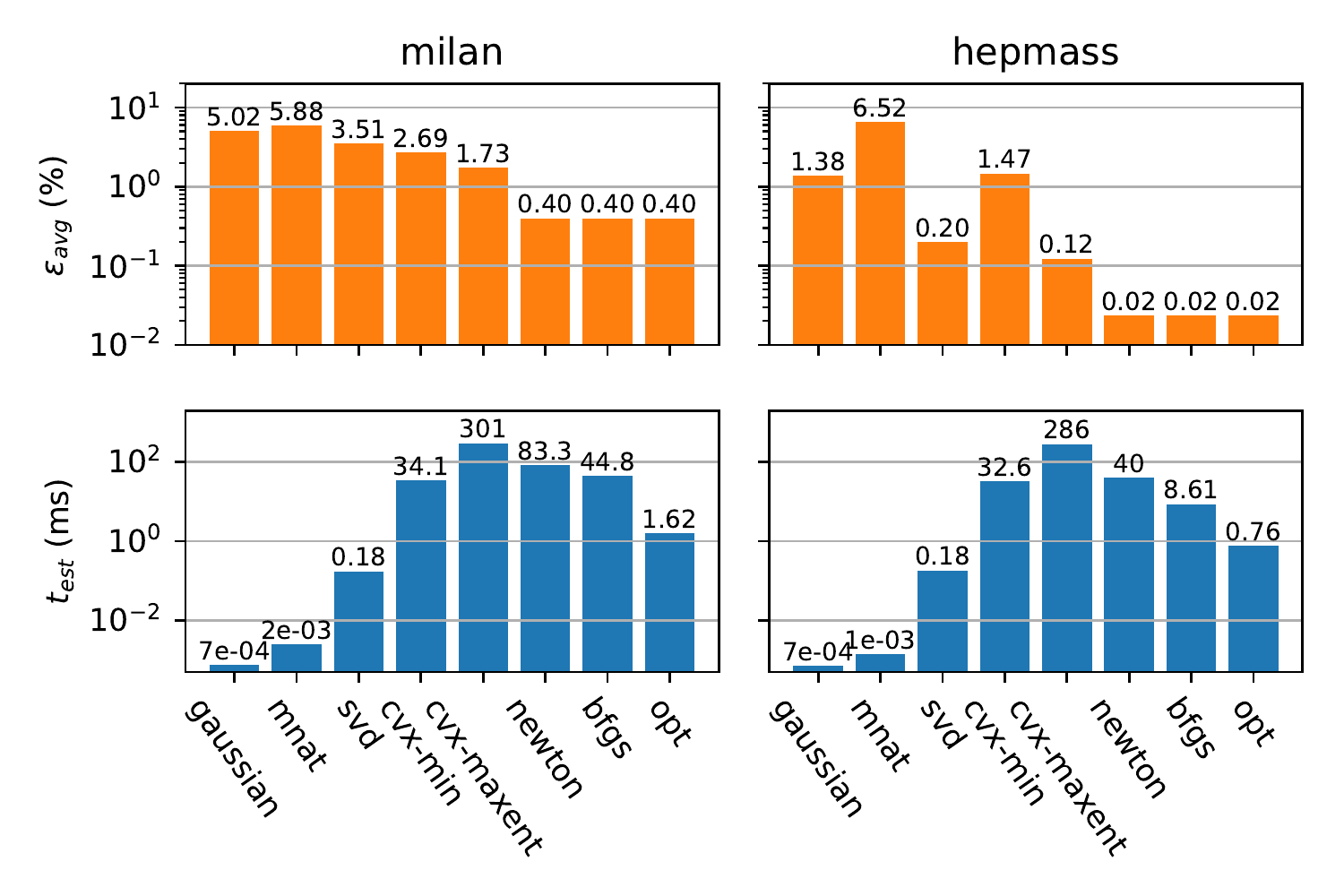}
    \vspace{-2em}
    \caption{Lesion study comparing our optimized maximum entropy solver to other estimators. Our \texttt{opt} estimator provides at least $5\times$ less error than estimators that do not use maximum entropy, and up to $200\times$ faster estimation times than naive maximum entropy solvers.}
    \label{fig:lesion}
    \vspace{-1em}
\end{figure}
Figure~\ref{fig:lesion} illustrates the average quantile error and estimation time for these estimators.
We run these experiments with $k=10$ moments.
For uniform comparisons with other estimators, on the milan dataset we only use the log moments, and on the hepmass dataset we only use the standard moments.
We perform discretizations using 1000 uniformly spaced points and make use of the ECOS convex solver~\cite{ecos}.
Solvers that use the maximum entropy principle provides at least $5\times$ less error than estimators that do not.
Furthermore, our optimizations are able to improve the estimation time by a factor of up to $200\times$ over an implementation using generic solvers, and provide faster solve times than naive Newton's method or BFGS optimizers.
As described in Section~\ref{sec:optimizations}, given the computations needed to calculate the gradient, one can compute the Hessian relatively cheaply, so our optimized Newton's method is faster than BFGS.

\section{Applying the Moments Sketch}
\label{sec:evalapps}
In this section, we evaluate how the \msketch affects performance when integrated with other data systems.
We examine how the \msketch improves query performance in the Druid analytics engine, as part of a cascade in the Macrobase feature selection engine~\cite{bailis2017macrobase}, and as part of exploratory sliding window queries. 

\subsection{Druid Integration}
\label{sec:eval_druid}
To illustrate the utility of the \msketch in a modern analytics engine, we integrate the \msketch with Druid~\cite{Yang2014druid}.
We do this by implementing \msketch as an user-defined aggregation extension, and compare the total query time on quantile queries using the \msketch with the default \stree summary used in Druid and introduced in~\cite{ben2010streaming}.
The authors in~\cite{ben2010streaming} observe on average 5\% error for an \stree with 100 centroids, so we benchmark a \msketch with $k=10$ against \strees with 10, 100, and 1000 centroids.

In our experiments, we deploy Druid on a single node -- the same machine described in section~\ref{sec:evalsetup} -- with the same base configuration used in the default Druid quickstart.
In particular, this configuration dedicates 2 threads to process aggregations.
Then, we ingest 26 million entries from the milan dataset at a one hour granularity and construct a cube over the grid ID and country dimensions, resulting in 10 million cells.

\begin{figure}
\includegraphics[width=\columnwidth]{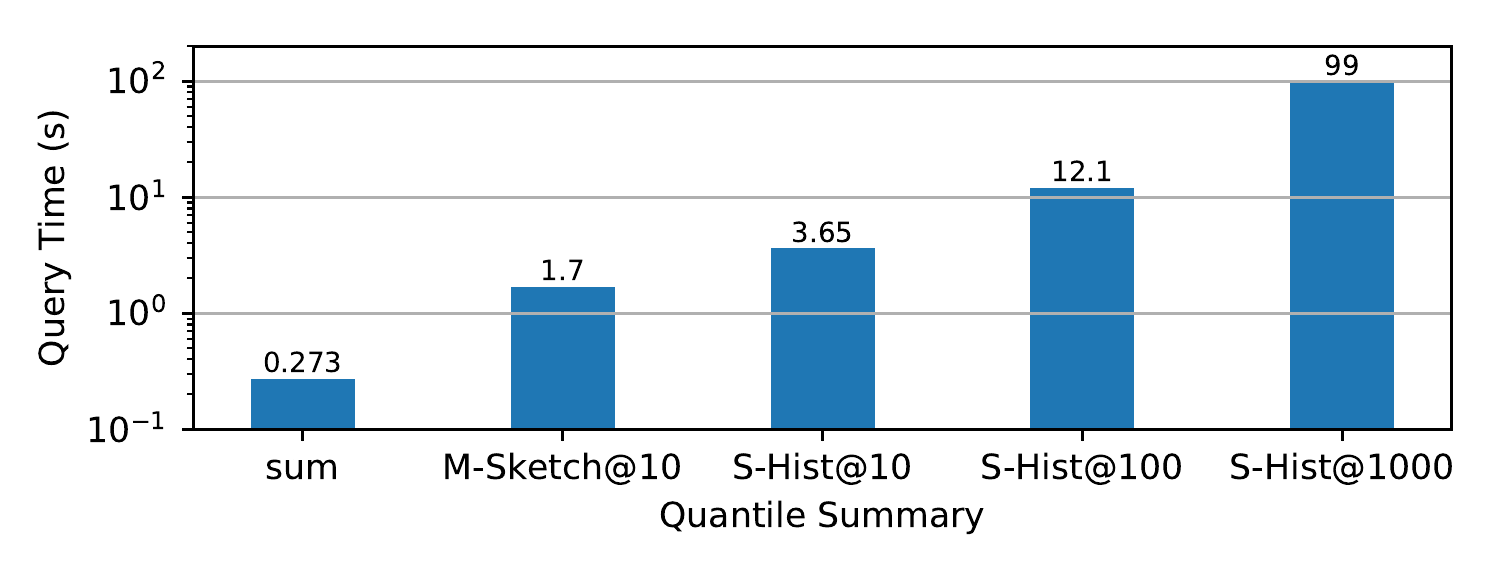}
\vspace{-2em}
\caption{Druid end-to-end query benchmark. The \msketch allows for faster query times than the comparable \stree summary with 100 bins. Runtime for a native sum operation is a lower bound on query time.}
\vspace{-1em}
\label{fig:druideval}
\end{figure}

Figure~\ref{fig:druideval} compares the total time to query for a quantile on the complete dataset using the different summaries.
The \msketch provides $7\times$ lower query times than a \stree with 100 bins. 
Furthermore, as discussed in Section~\ref{sec:evalquerytime}, any \stree with fewer than 10 thousand buckets provides worse accuracy on milan data than the \msketch.
As a best-case baseline, we also show the time taken to compute a native sum query on the same data.
The 1 ms cost of solving for quantile estimates from the \msketch on this dataset is negligible here.
\label{sec:sketch}

\subsection{Threshold queries}
\label{sec:eval_threshold}
In this section we evaluate how the cascades described in Section~\ref{sec:cascades} improve performance on threshold predicates.
First we show in Section~\ref{sec:eval_mb} how the MacroBase analytics engine can use the \msketch to search for anomalous dimension values.
Then, we show in Section~\ref{sec:eval_sliding} how historical analytics queries can use the \msketch to search and alert on sliding windows.

\subsubsection{MacroBase Integration}
\label{sec:eval_mb}
The MacroBase engine searches for dimension values with unusually high outlier rates in a dataset \cite{bailis2017macrobase}.
For example, given an overall 2\% outlier rate, MacroBase may report when a specific app version has an outlier rate of 20\%.
We integrate the \msketch with a simplified deployment of MacroBase where all values greater than the global 99th percentile $t_{99}$ are considered outliers.
We then query MacroBase for all dimension values with outlier rate at least $r=30\times$ greater than the overall outlier rate.
This is equivalent to finding subpopulations whose 70th percentile is greater than $t_{99}$.

Given a cube with pre-aggregated \msketches for each dimension value combination and no materialized roll-ups, 
MacroBase merges the \msketches to calculate the global $t_{99}$, and then runs Algorithm~\ref{alg:cascade} on every dimension-value subpopulation, searching for subgroups with $q_{.7} > t_{99}$.
We evaluate the performance of this query on 80 million rows of the milan internet usage data from November 2013, pre-aggregated by grid ID, country, and at a four hour granularity.
This resulted in 13 million cube cells, each with its own \msketch.

Running the MacroBase query produces 19 candidate dimension values.
We compare the total time to process this query using direct quantile estimates, our cascades, and the alternative \yahoo quantile sketch.
In the first approach (\texttt{Merge12a}), we merge summaries during MacroBase execution as we do with a \msketch.
In the second approach (\texttt{Merge12b}), we calculate the number of values greater than the $t_{99}$ for each dimension value combination and accumulate these counts directly, instead of the sketches.
We present this as an optimistic baseline, and is not always a feasible substitute for merging summaries.

\begin{figure}
\includegraphics[width=\columnwidth]{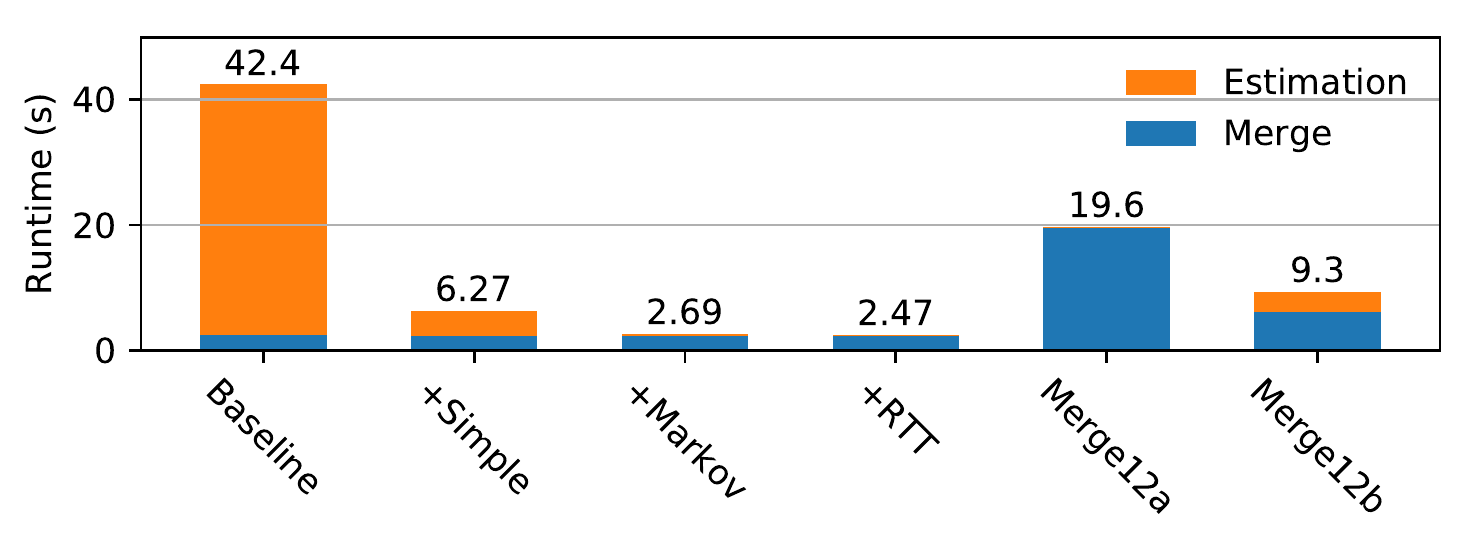}
\vspace{-2em}
\caption{Runtime of MacroBase queries: the final \msketch cascade outperforms queries using alternate sketches.}
\vspace{-1em}
\label{fig:cascadeeval2}
\end{figure}
Figure~\ref{fig:cascadeeval2} shows the query times for these different methods: the baseline method calculates quantile estimates directly, we show the effect of incrementally adding each stage of our cascade ending with +\racz.
Each successive stage of the cascade improves query time substantially.
With the complete cascade, estimation time is negligible compared to merge time.
Furthermore, the \msketch with cascades has $7.9\times$ lower query times than using the \yahoo sketch, and even $3.7\times$ lower query times than the \texttt{Merge12b} baseline.

\begin{figure}
\includegraphics[width=\columnwidth]{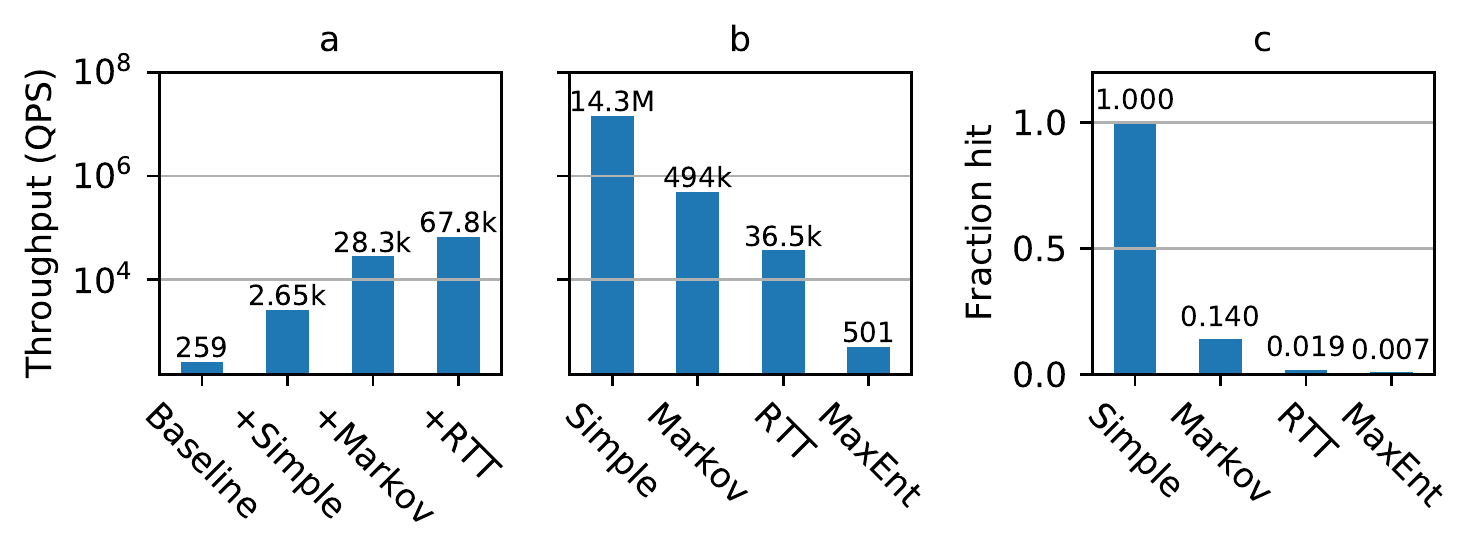}
\vspace{-2em}
\caption{Cascades in MacroBase: (a) as we incrementally add cascade stages, threshold query throughput increases. (b) The cascade proceeds from faster to slower estimates. (c) Each stage of the cascade processes a smaller fraction of queries.}
\vspace{-1em}
\label{fig:cascadeeval1}
\end{figure}
In Figure~\ref{fig:cascadeeval1} we examine the impact the cascade has on estimation time directly. 
Each additional cascade stage improves threshold query throughput and is more expensive than the last.
The complete cascade is over 250$\times$ faster than this baseline, and $25\times$ faster than just using a simple range check.

\subsubsection{Sliding Window Queries}
\label{sec:eval_sliding}
Threshold predicates are broadly applicable in data exploration queries.
In this section, we evaluate how the \msketch performs on sliding window alerting queries.
This is useful when, for instance, users are searching for time windows of unusually high CPU usage spikes.

For this benchmark, we aggregated the 80 million rows of the milan dataset at a 10-minute granularity, which produced 4320 panes that spanned the month of November.
We augmented the milan data with two spikes corresponding to hypothetical anomalies.
Each spike spanned a two-hour time frame and contributed 10\% more data to those time frames.
Given a global 99th percentile of around 500 and a maximum value of 8000, we added spikes with values $x=2000$ and $x=1000$

We then queried for the 4-hour time windows whose 99th percentile was above a threshold $t=1500$.
When processing this query using a \msketch, we can update sliding windows using turnstile semantics, subtracting the values from the oldest pane and merging in the new one, and use our cascade to filter windows with quantiles above the threshold.

\begin{figure}
\includegraphics[width=\columnwidth]{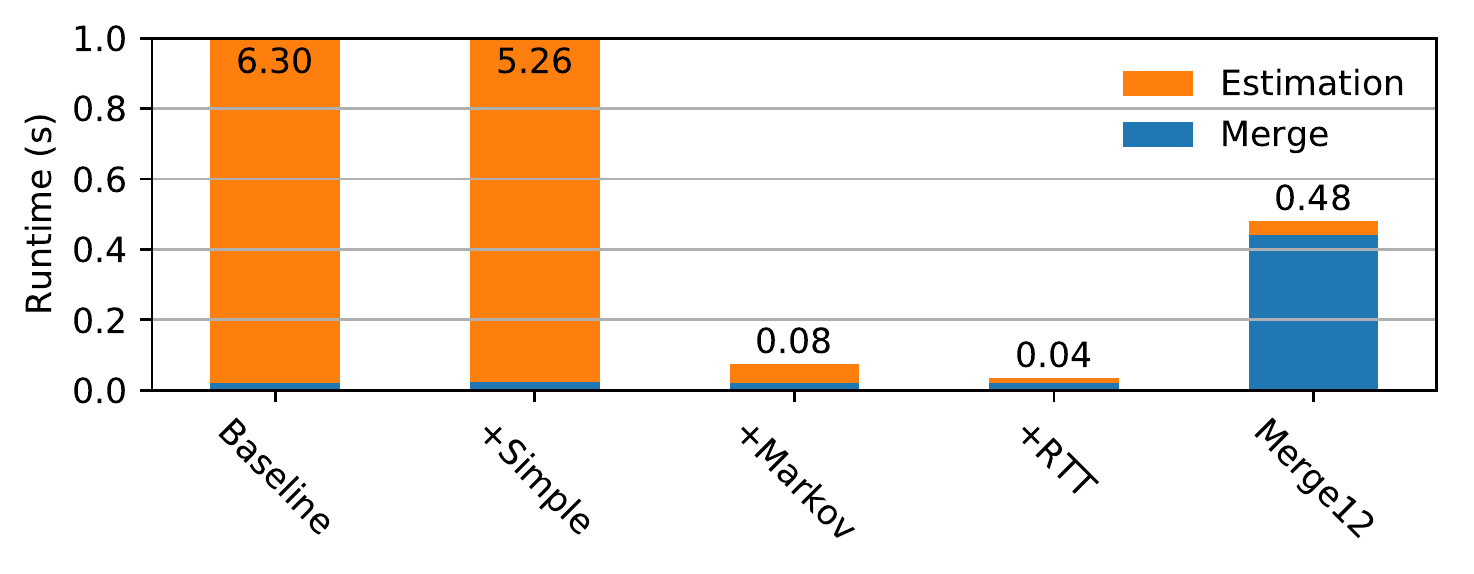}
\vspace{-2em}
\caption{Sliding window query: \msketch with cascades runs 13$\times$ faster than $\texttt{Merge12}$.}
\vspace{-1em}
\label{fig:slidingwindow}
\end{figure}
Figure~\ref{fig:slidingwindow} shows the runtime of the sliding window query using both the \msketch and \texttt{Merge12}.
Faster \msketch merge times and the use of turnstile semantics then allow for 13$\times$ faster queries than \texttt{Merge12}.
\section{Conclusion}
\label{sec:conclusion}
In this paper, we show how to improve the performance of quantile aggregation queries using statistical moments.
Low merge overhead allows the \msketch to outperform comparably accurate existing summaries when queries aggregate more than 10 thousand summaries.
By making use of the method of moments and the maximum entropy principle, the \msketch provides $\epsilon_\mathrm{avg} \leq 0.01$ accuracy on real-world datasets, while the use of numeric optimizations and cascades keep query times at interactive latencies.

{
\section*{Acknowledgments}
This research was made possible with feedback and assistance from our collaborators at Microsoft including Atul Shenoy, Will Mortl, Cristian Grajdeanu, Asvin Ananthanarayan, and John Sheu. This research was supported in part by affiliate members and other supporters of the Stanford DAWN project -- Facebook, Google, Intel, Microsoft, NEC, Teradata, VMware, and SAP -- as well as Toyota, Keysight Technologies, Hitachi, Northrop Grumman, Amazon Web Services, Juniper, NetApp, and the NSF under CAREER grant CNS-1651570 and GRFP grant DGE-114747.
}

\iftoggle{arxiv}{
\bibliographystyle{ACM-Reference-Format}
\Urlmuskip=0mu plus 1mu
\bibliography{moments} 
}{
{
\small
\bibliographystyle{abbrv}
\Urlmuskip=0mu plus 1mu
\bibliography{moments} 
}
}

\clearpage
\appendix

\section{Maximum Entropy Estimation}
\label{sec:appendix_maxent}

\subsection{Newton's Method Details}
Recall that we wish to solve the following optimization problem:
\begin{align*}
  \underset{f\in\mathcal{F}[x_\mathrm{min}, x_\mathrm{max}]}{\mathrm{maximize}}& H[f] \\
  \text{subject to } & \int_{x_\mathrm{min}}^{x_\mathrm{max}} x^i f(x)\,dx = \mu_i, & i \in \{1, \dots, k_1\}  \\
    & \int_{x_\mathrm{min}}^{x_\mathrm{max}} \log^i(x) f(x) \, dx = \nu_i, & i \in \{1, \dots, k_2\} 
\end{align*}.

Throughout this section it is easier to reformulate this problem in more general terms, using the functions
\begin{align}
m_i(x) =  
\begin{cases}
    x^i & 0 \leq i \leq k_1 \\
    h(x)^{i-k_1} & k_1+1 \leq i \leq k_1+k_2.
\end{cases}
\label{eqn:defm}
\end{align}
where $h(x) = \log(x)$ or $h(x) = e^{x}$ depending on whether we work using the $x$ or log-transformed $x'=\log(x)$ as our primary metric.
Letting $k_t = k_1+k_2$, and folding the $\nu_i$ into a larger $\vec{\mu}$ vector, our optimization problem is then:
\begin{align}
  \underset{f\in\mathcal{F}[x_\mathrm{min}, x_\mathrm{max}]}{\mathrm{maximize}}& H[f] \label{eqn:genoptprob}\\
  \text{subject to } & \int_{x_\mathrm{min}}^{x_\mathrm{max}} m_i(x) f(x)\,dx = \mu_i, & i \in \{0, \dots, k_t\} \nonumber
\end{align}.

Functional analysis \cite{jaynes1957} tells us that a maximal entropy solution to Eq.~\eqref{eqn:genoptprob} has the form:
\begin{equation*}
  f(x;\theta) = \exp\left(\sum_{i=0}^{k_t} \theta_i m_i(x)\right),
\end{equation*}

Then if we define the potential function $L(\theta)$ from~\cite{mead1984maxent}:
\begin{align}
L(\theta) =& \int_{x_\mathrm{min}}^{x_\mathrm{min}} \exp\left(\sum_{i=0}^{k_t} \theta_i m_i(x)\right) - \sum_{i=0}^{k_t} \theta_i \mu_i \label{eqn:genpotential}
\end{align}

We can calculate the gradient and Hessian of $L(\theta)$ as follows:
\begin{equation}
\frac{\partial L}{\partial \theta_i} = \int_{x_\mathrm{min}}^{x_\mathrm{min}} m_i(x) \exp{\left(\sum_{i=0}^{k_t} \theta_i m_i(x) \right)} - \mu_i \label{eqn:grad} 
\end{equation}
\begin{equation}
\frac{\partial^2 \Gamma}{\partial \theta_i \partial \theta_j} = \int_{x_\mathrm{min}}^{x_\mathrm{min}} m_i(x)m_j(x) \exp{\left(\sum_{i=0}^{k_t} \theta_i m_i(x) \right)} \label{eqn:hessiant}
\end{equation}

Note that when the gradient given in Eq.~\eqref{eqn:grad} is zero then the constraints in Eq.~\eqref{eqn:genoptprob} are satisfied.
Since $L(\theta)$ is convex and has domain $\mathbb{R}^{k_d}$, this means that by solving the unconstrained minimization problem over $L(\theta)$ we can find a solution $\theta$ we can find a solution to the constrained maximum entropy problem.
Since Newton's method is a second order method,
we can use Equations \eqref{eqn:genpotential}, \eqref{eqn:grad}, \eqref{eqn:hessiant} are to execute Newton's method with backtracking line search \cite{boydcvx}.

\subsection{Practical Implementation Details}
\minihead{Chebyshev Polynomial Basis Functions}

As described in Section~\ref{sec:optimizations}, we can improve the stability of our optimization problem by using Chebyshev polynomials.
To do so, we must redefine our $m_i(x)$
\begin{equation*}
    m_i(x) = 
    \begin{cases}
        T_i(s_1(x)), &i \in\{1, \dots, k_1\} \\
        T_{i-k_1}(s_2(h(x))), &i \in\{k_1 + 1, \dots, k_1 + k_2\}
    \end{cases}
\end{equation*}
where $T_i(x)$ are Chebyshev polynomials of the first kind~\cite{press2007numericalrecipes} and the $s_1,s_2$ are linear scaling functions to map onto $[-1,1]$ defined as:
\begin{align*}
s_1(x) &= \left(x - \frac{x_\mathrm{max}+x_\mathrm{min}}{2}\right)/\left(\frac{x_\mathrm{max}-x_\mathrm{min}}{2}\right) \\
s_2(x) &= \left(x - \frac{h(x_\mathrm{max})+h(x_\mathrm{min})}{2}\right)/\left(\frac{h(x_\mathrm{max})-h(x_\mathrm{min})}{2}\right)
.
\end{align*}

The formulae for $\mu_i, \Gamma, \nabla \Gamma, \nabla^2 \Gamma$ still hold, but now the $\mu_i$ are 
\begin{equation}
    \mu_i = \frac{1}{n}
    \begin{cases}
        \sum_x T_i(s_1(x)) & 0 \leq i \leq k_1 \\
        \sum_x T_{i-k_a}(s_2(h(x))) & k_1+1 \leq i \leq k_1+k_2.
    \end{cases}.
\end{equation}
These can be computed from the quantities $\mu_i = \sum_x x^i$, $\nu_i = \sum_x h(x)^i$ originally stored in the \msketch by using the binomial expansion and standard formulae for expressing Chebyshev polynomials in terms of standard monomials \cite{mason2002chebyshev}.

\minihead{Chebyshev Polynomial Integration}

In this section we will show how Chebyshev approximation provides for efficient ways to compute the gradient and Hessian.
Here it is easier to work with the change of variables $u=s_1(x)$ so that $f^u(u) = f(s_1^{-1}(u))$ has domain $[-1,1]$.
First, we will examine the case when $k_2 = 0$.
If we can approximate $f(u;\theta)$ as a linear combination of chebyshev polynomials:
\begin{align}
  f^u(u;\theta) &= \exp\left(\sum_{i=0}^{k_t} \theta_i m_i(u)\right) \\
  &\approx \sum_{j=0}^{n_c} c_j T_j(u) \label{eqn:fapprox}
\end{align}

Then using polynomial identities such as $T_i(x)T_j(x) = \frac{1}{2}(T_{i+j}(x) + T_{|i-j|}(x))$ we can evaluate the Gradient and Hessian in Eqs. \eqref{eqn:grad}, \eqref{eqn:hessiant} using $O(k_1 n_c)$ algebraic operations.

We can approximate $f^u(u;\theta)$ using the Clenshaw Curtis quadrature formulas to approximate a function $f$ supported on $[-1,1]$ (Eq. 5.9.4 in \cite{press2007numericalrecipes}):
\begin{equation}
a_j = \frac{2}{n_c}\left(\frac{f(1)}{2} - \frac{f(-1)}{2} + \sum_{i=1}^{n_c-1} f\left[\cos\left(\frac{\pi i}{n_c}\right)\right]\cos\left(\frac{\pi i}{n_c}\right)\right)
\label{eqn:chebyapproxcoeff}
\end{equation}
Then 
\begin{equation}
f(x) \approx \frac{1}{2}T_0(x) + \sum_{i=1}^{n_c} a_i T_i(x)
\end{equation}
where Eq.~\eqref{eqn:chebyapproxcoeff} can be evaluated in $n_c\log{n_c}$ time using the Fact Cosine Transform~\cite{press2007numericalrecipes}.
The case when $k_2 > 0$ is similar except we need to approximate not just $f^u(u;\theta)$ but also $T_i(s_2(h(s_1^{-1}(u)))f^u(u;\theta)$ for $i\leq k_2$.

\section{Numeric Stability of Moments}
\label{sec:appendix_stability}
As mentioned in Section~\ref{sec:choosek}, floating point arithmetic limits the usefulness of higher moments.
This is because the raw moments $\frac{1}{n}\sum x^i$ are difficult to optimize over and analyze: both maximum entropy estimation (Section~\ref{sec:optimizations}) and theoretical error bounds (Section~\ref{sec:errorgeneral}) apply naturally to moments on data in the range $[-1,1]$.
In particular, shifting the data improves the conditoning of the optimization problem dramatically.
However, when merging sketches from different datasets, users may not know the full range of the data ahead of time, so the power sums stored in a \msketch correspond to data in an arbitrary range $[a,b]$.
Thus, we will analyze how floating point arithmetic affects the process of scaling and shifting the data so that it is supported in the range $[-1,1]$.
This is similar to the process of calculating a variance by shifting the data so that it is centered at zero.

We can calculate the moments of scaling data $x_{scale}=k\cdot x$ with error only in the last digit, so we can assume we have data in the range $[c-1,c+1]$. 
Let $\mu_i$ be the moments of the $x_{scale}$, and let $\mu_i^s$ be the moments of the shifted data $x_{shift} = x_{scale} - c$.
Using the binomial expansion 
\begin{align*}
\mu_k^s &= \frac{1}{n}\sum_{x \in x_{scale}} \left(x - c\right)^k \\
    &= \sum_{i=0}^{k} {k \choose i} \mu_i (-c)^{k-i}
\end{align*}
Using numerically stable addition, we can calculate $\mu_k$ to a relative precision $\delta$ close to machine precision $\delta \approx 2^{-53}$.
Then, the absolute error $\delta_k$ in estimating $\mu_k^s$ is bounded by:
\begin{align*}
\delta_k &\leq \sum_{i=0}^{k} {k \choose i} |\mu_i| |c|^{k-i} \delta \\
\delta_k &\leq \sum_{i=0}^{k} {k \choose i} (|c|+1)^k \delta \\
& \leq 2^k (|c|+1)^k \delta
\end{align*}
We know that the average quantile error (Equation~\ref{eqn:avgerrorrange} in Section~\ref{sec:errorgeneral}) is bounded by 
$$\epsilon_{\textrm{avg}} \leq O\left(\frac{1}{k} + 3^k \|\mu_f - \mu_{\hat{f}}\|_2\right),$$ 
so if we can calculate all of the $\mu_i^s$ to within precision $3^{-k}\left(\frac{1}{k-1} - \frac{1}{k}\right)$ then we have enough precision to bound the quantile estimate by $O\left(\frac{1}{k-1}\right)$.
This way, we can show that the error bound from using the first $k$ moments will be at least as tight as the bound from using the first $k-1$ moments.
As $k$ and $|c|$ grow, achieving this precision becomes more and more difficult, and we can solve for the cutoff point using base-10 $\log$.
\begin{align}
2^k (|c|+1)^k \delta & \leq 3^{-k}\left(\frac{1}{k-1} - \frac{1}{k}\right) \\
k\left(\log{6} + \log{(|c|+1)}\right) & \leq \log{\frac{1}{\delta}} - \log{(k^2 - k)}) \label{eqn:krawbound}
\end{align}

Plugging in double precision for $\delta$ into Eq.~\eqref{eqn:krawbound}, we know that $k\leq \frac{53\log{2}}{\log{6}} \leq 20$, so $\log{(k^2 - k)} \leq 2.58$
\begin{align}
k &\leq \frac{53\log{2} - 2.58}{\log{6} + \log{(|c| + 1)}} \\
 &\leq \frac{13.35}{.78 + \log{(|c|+1)}}
\label{eqn:maxkformula}
\end{align}

\begin{figure}
    \includegraphics[width=\columnwidth]{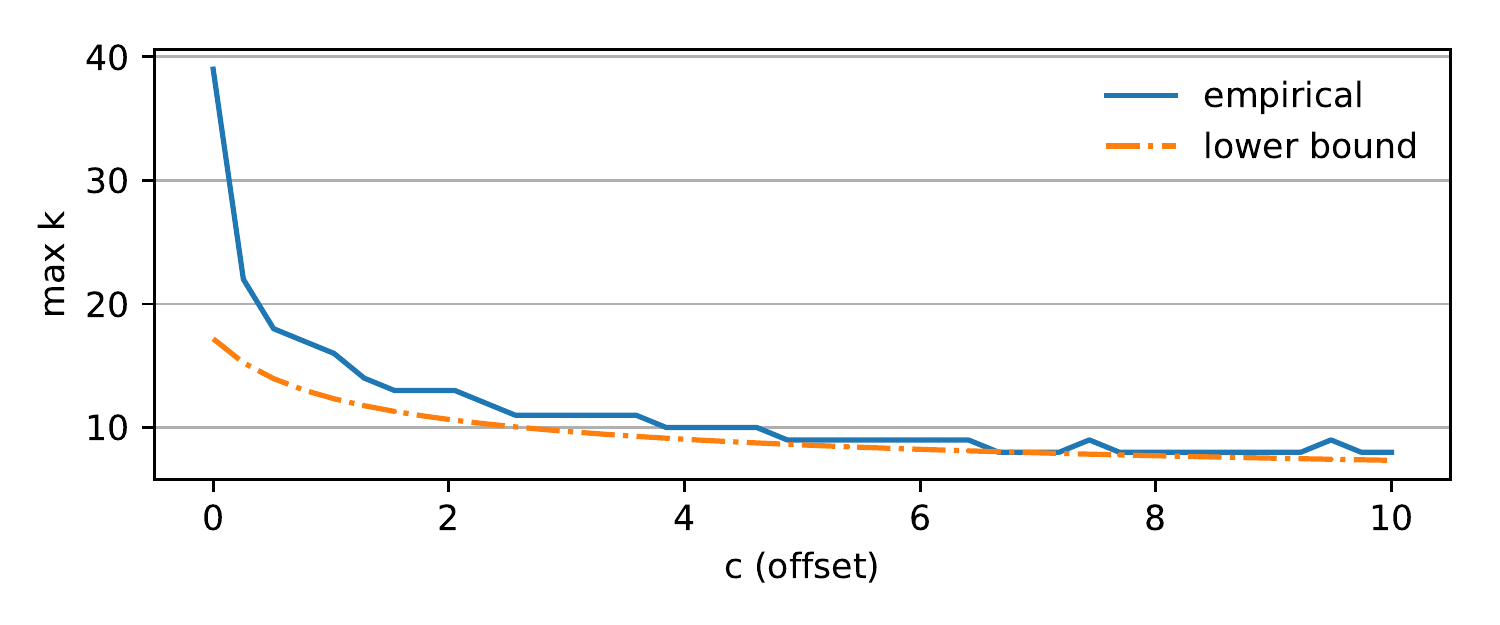}
    \vspace{-2em}
    \caption{Highest order usable moments for data centered at different locations. Our data-independent bound is conservative compared to values on a uniform dataset.}
    \vspace{-1em}
    \label{fig:max_moments}
\end{figure}

Equation~\ref{eqn:maxkformula} is a conservative bound on the number of numerically stable moments we can extract from an \msketch, and suggests that when our data is centered at 0, we have at least $17$ stable moments.
When the raw data have range $[x_{\mathrm{min}}, 3x_{\mathrm{min}}]$, then $c=2$, and we have at least $10$ stable moments.
In our evaluations, $10$ stable moments are enough to achieve quantile error $\approx .01$.
Figure~\ref{fig:max_moments} describes how the bound in Equation~\ref{eqn:maxkformula} varies with $c$, and compares it with the highest order stable moment of a uniform distribution supported on $[c-1,c+1]$.
This confirms that our formula is a conservative bound on the true precision loss due to the shift.
If the raw data are centered even further from 0, users can consider pre-shifting all of their data to make better use of numeric precision.

\begin{figure}
    \includegraphics[width=\columnwidth]{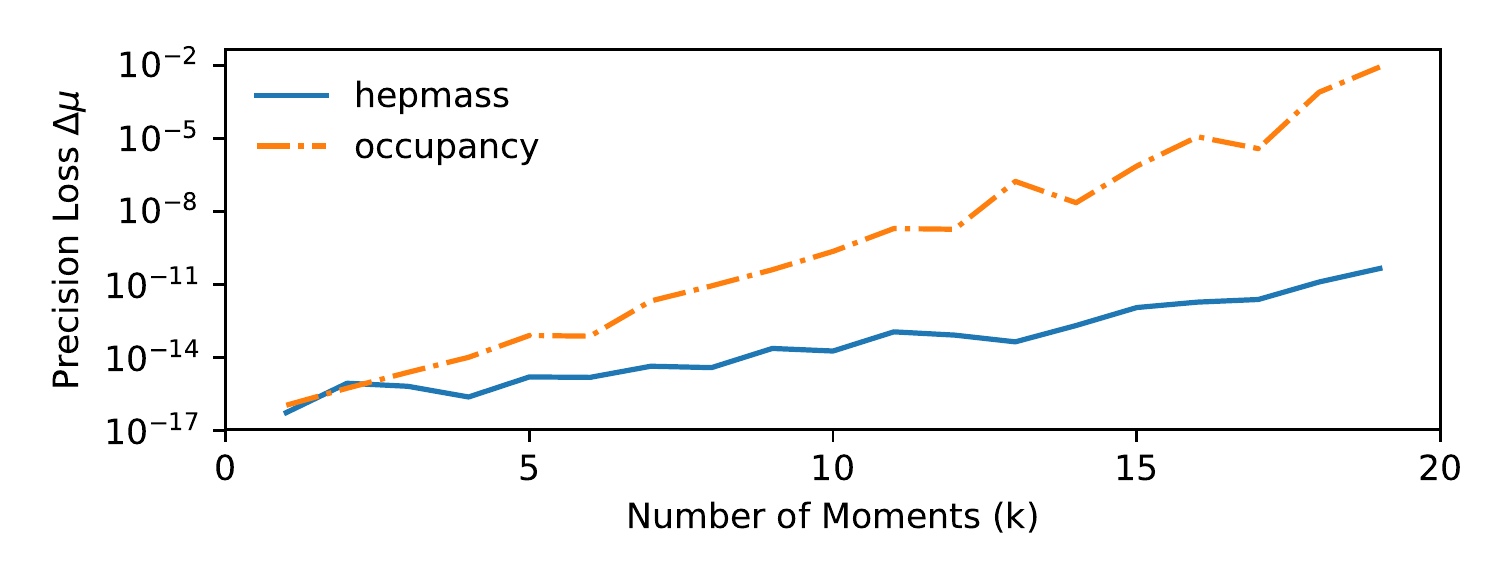}
    \vspace{-2em}
    \caption{Precision loss from shifting and converting higher moments to chebyshev moments. The occupancy dataset exhibits more precision loss because it is centered further away from zero.}
    \vspace{-1em}
    \label{fig:cheby_precision}
\end{figure}

As a measure of the downstream impact of this effect on some of our evaluation datasets, Figure~\ref{fig:cheby_precision} shows the precision loss during Chebyshev polynomial calculation $\Delta \mu = |\mu_i - \hat\mu_i|$ where $\mu_i$ is the true Chebyshev moment and $\hat\mu_i$ is the value calculated from the \msketch.
Precision loss is more severe on the occupancy dataset which is centered away from zero ($c\approx 1.5$) compared with the hepmass dataset ($c \approx 0.4$). 
See Table~\ref{tab:data_params}.

\section{Low-precision storage}
\label{sec:eval_lowprecision}
In Appendix~\ref{sec:appendix_stability} we discussed how floating point precision limits the usability of higher moments.
Conversely, in settings where space is heavily constrained, the data is well-centered, and we only need a limited number of moments, the \msketch can be compressed by reducing the precision of the sketch contents using randomized rounding.

As a proof-of-concept of this approach, we created an encoder that compresses the double precision floating point values in a \msketch using reduced floating point precision, quantizing the significand and removing unused bits in the exponent.
This low-precision representation has a negligible impact on merge times since we can convert them to and from native double precision using simple bit manipulation.

\begin{figure}
\includegraphics[width=\columnwidth]{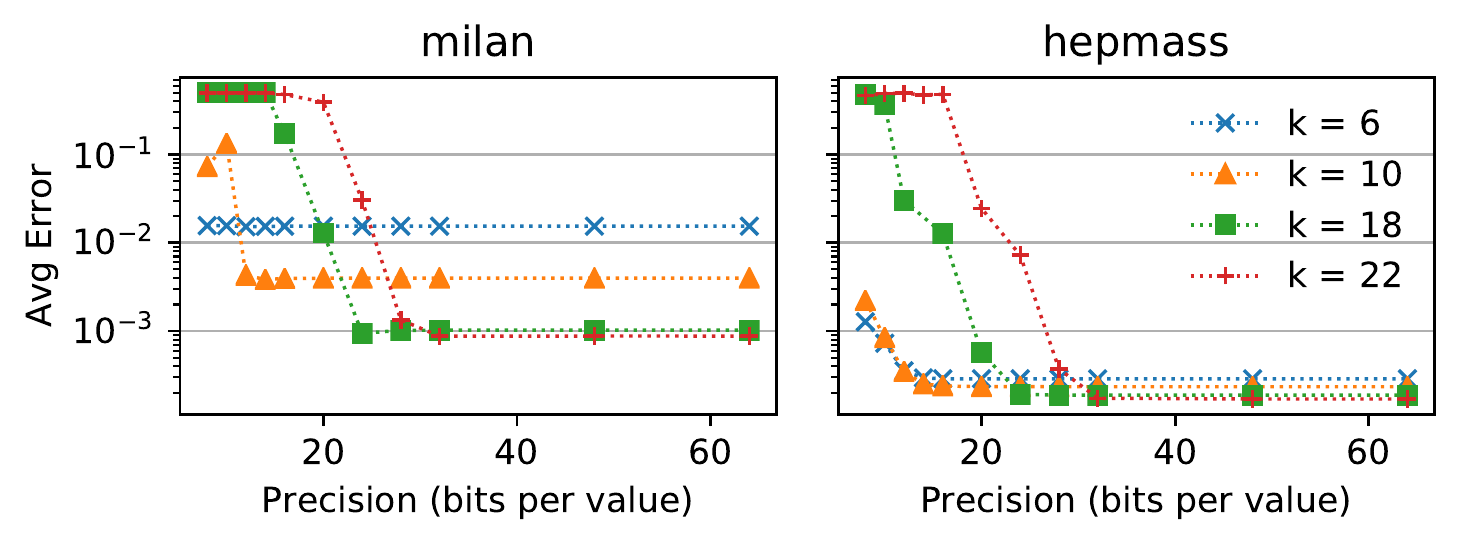}
\vspace{-2em}
\caption{Average error for low-precision \msketches after 100 thousand merges. Twenty bits of precision is sufficient to maintain accuracy for both datasets.}
\vspace{-1em}
\label{fig:lowprecision}
\end{figure}

We evaluate the encoding by constructing 100 thousand pre-aggregated \msketches, reducing their precision, and then merging them and querying for quantiles on the aggregation.
Figure~\ref{fig:lowprecision} illustrates how the quality of the final estimate remains stable as the precision is decreased until we reach a minimum threshold, after which accuracy degrades.
On the milan dataset, a \msketch with $k=10$ can be stored with 20 bits per value without noticeably affecting our quantile estimates, representing a $3\times$ space reduction compared to standard double precision floating point.

These results are consistent with the bounds in Section~\ref{sec:appendix_stability} and show how higher moments require more bits of precision.
However, the bounds are conservative since they only consider the midpoint $c$ of a dataset and are otherwise both dataset-independent and agnostic to the maximum entropy principle.

\section{Additional Workloads}
\label{sec:varydistribution}

The \msketch accuracy and performance generalizes across a range of workloads. In this section we evaluate its performance on datasets with varying skew, in the presence of outlier values, under a coarser pre-aggregation policy, and on a production workload.

\subsection{Data Skew}
\begin{figure}
    \includegraphics[width=\columnwidth]{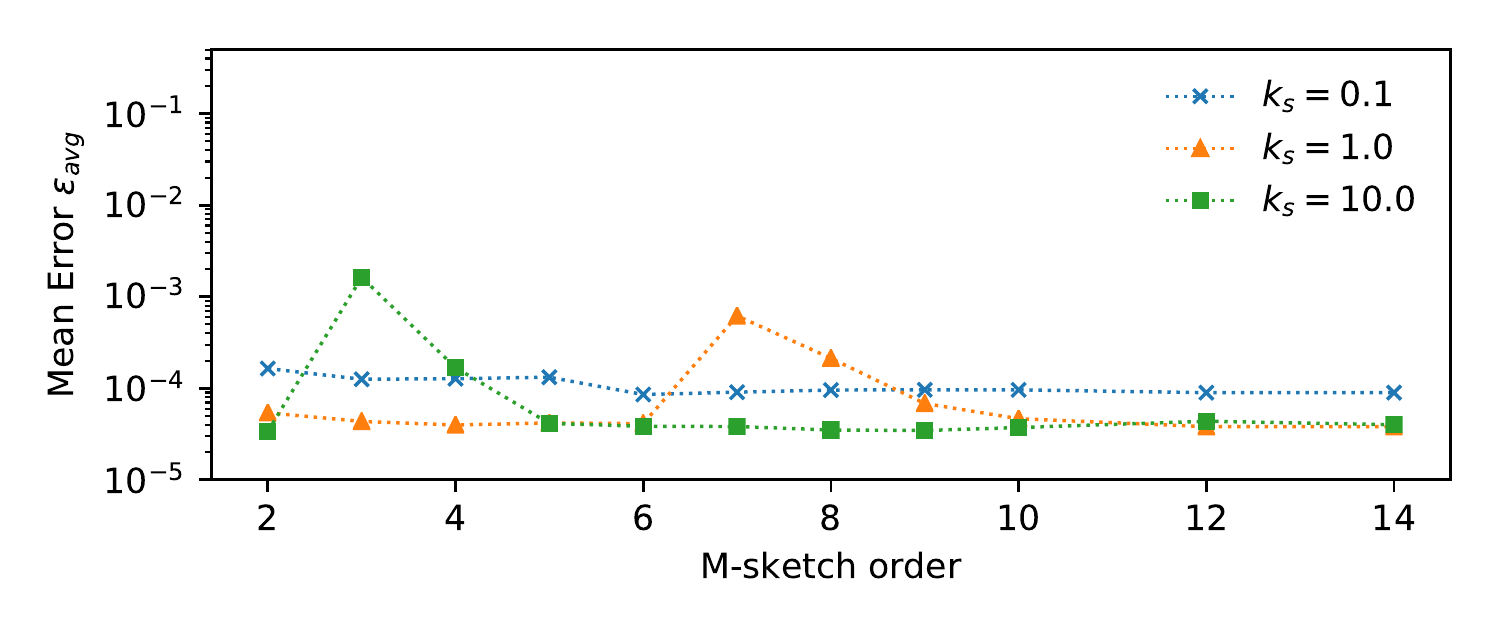}
    \vspace{-2em}
    \caption{Accuracy of estimates on Gamma distributions with varying shape parameter $k_{s}$. The maximum entropy principle is able to construct an accurate estimate across a range of parameters.}
    \label{fig:varyskew}
    \vspace{-1em}
\end{figure}

Our usage of log-moments greatly reduces the impact of data skew on the accuracy of \msketch quantile estimates.
In Figure~\ref{fig:varyskew} we vary the shape parameter $k_{s}$ of a Gamma distribution with scale factor $\theta=1$.
The skew of this distribution is $\frac{2}{\sqrt{k_{s}}}$ so $k_{s}=0.1$ corresponds to very high skew.
For $k_{s}=0.1,1.0,10.0$, our estimator can achieve $\epsilon_{avg}\leq 10^{-3}$ error.
The accuracy regressions on orders $3$ and $7$ occur when the solver stops making use of all available moments to reduce the condition number of the Hessian (Section~\ref{sec:optimizations}).
In this specific case, our solver uses a heuristic to decide that given a maximum condition number, optimizing using 3 log moments is more valuable than 2 log moments and 2 standard moments.
This choice leads to worse accuracy on a Gamma distribution, but in general it is difficult to know which subset of moments will yield the most accurate estimate. 
More effective heuristics for choosing subsets of moments that do not exceed a condition number threshold is an open direction for future research.

\subsection{Outlier Values}
\begin{figure}
    \includegraphics[width=\columnwidth]{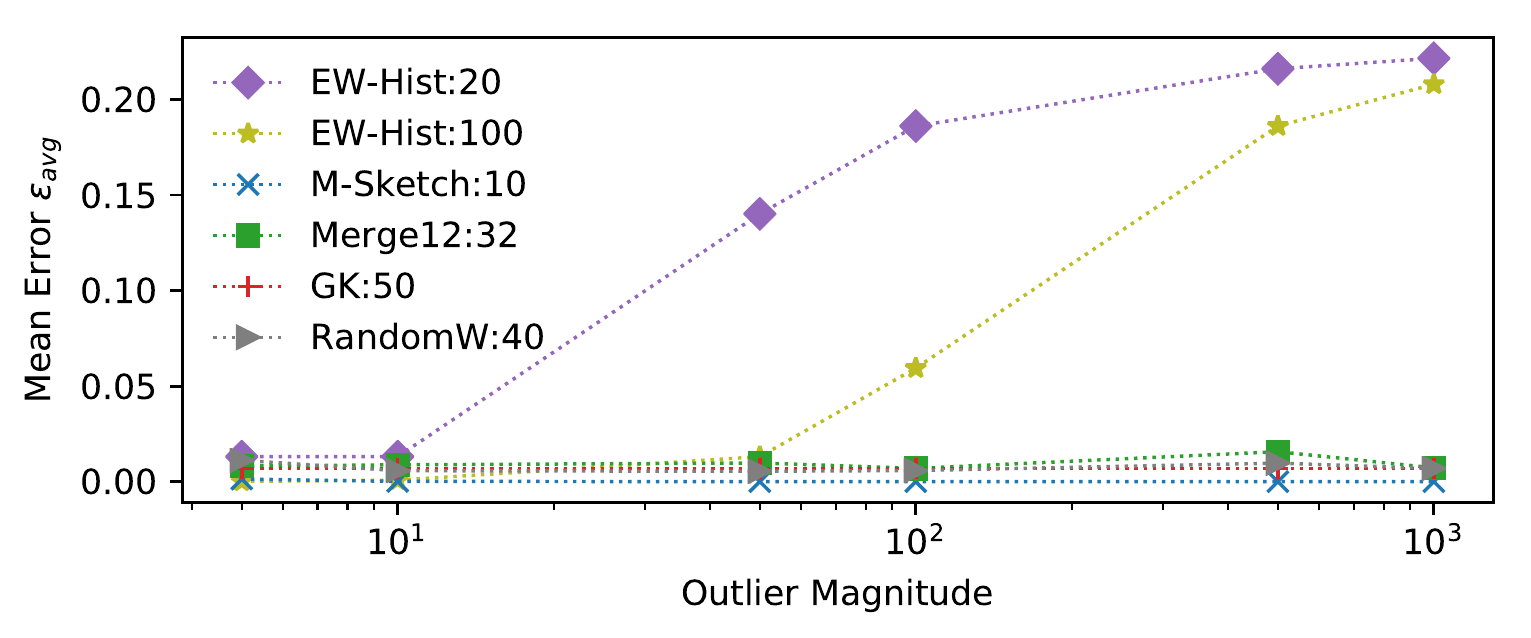}
    \vspace{-2em}
    \caption{Mean error on a Gaussian dataset with outliers of different magnitudes added. The \msketch remains accurate for large outliers, but the \hist accuracy degrades.}
    \vspace{-1em}
    \label{fig:robust_outlier}
\end{figure}

The \msketch, unlike histogram-based sketches, is also somewhat robust to the presence of large outlier values in a dataset.
In Figure~\ref{fig:robust_outlier} we evaluate the effect of adding a fixed fraction $\delta=0.01$ of outlier values from a Gaussian with mean $\mu_o$ and standard deviation $\sigma=0.1$ to a dataset of 10 million standard Gaussian points.
As we increase the magnitude $\mu_o$ of the outliers, the \hist summaries with 20 and 100 bins lose accuracy though a \msketch with $k=10$ remains accurate.
The \yahoo sketch is agnostic to value magnitudes and is unaffected by the outliers.
If extremely large outliers are expected, floating point precision suffers and the \msketch can be used in conjunction with standard outlier removal techniques.

\subsection{Varying Aggregation}
\label{sec:appendix_cellsize}
In our main evaluations, we group our datasets into cells of 200 elements and construct sketches for each cell to maintain a pre-aggregated collection of data summaries.
We do not target deployments where very few elements can be pre-aggregated per summary: in these cases merging \msketches is relatively expensive.
On the other hand production data systems can have much larger data volumes and opportunities to pre-aggregate more elements per cell.
Since the \msketch is fixed-size regardless of the data, increasing the number of elements per cell does not affect its merge time performance, while other sketches which which have not reached their maximum capacity will be correspondingly larger and slower to merge.

\begin{figure}
    \includegraphics[width=\columnwidth]{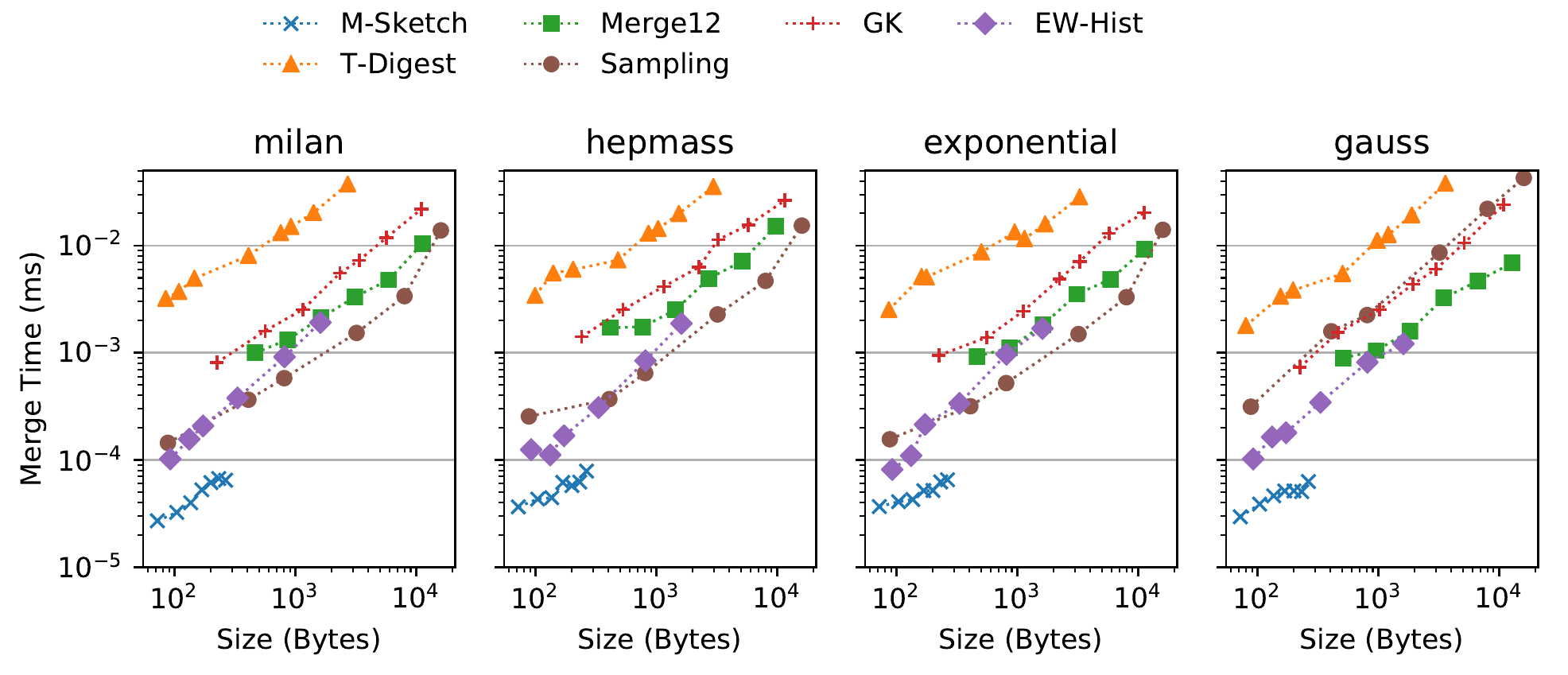}
    \vspace{-2em}
    \caption{Merge times on with sketches on cells of 2000 elements, and on a Gaussian dataset with cells of 10000 elements. Since the \msketch has a fixed size, its per-merge times remain faster than alternative sketches with comparable accuracy.}
    \label{fig:mergebig}
\end{figure}

In Figure~\ref{fig:mergebig} we measure the time taken per merge for different summaries constructed on cells of 2000 elements for the milan, hepmass, and exponential dataset, and cells of 10000 elements on a synthetic Gaussian dataset with 1 billion points.
The relative performance of different sketches matches closely with Figure~\ref{fig:mergetime}, except that larger \sampling and \yahoo summaries are now slower when constructed on more than 200 elements.

\subsection{Production workload}
\label{sec:appendix_aria}
\begin{figure}
    \includegraphics[width=\columnwidth]{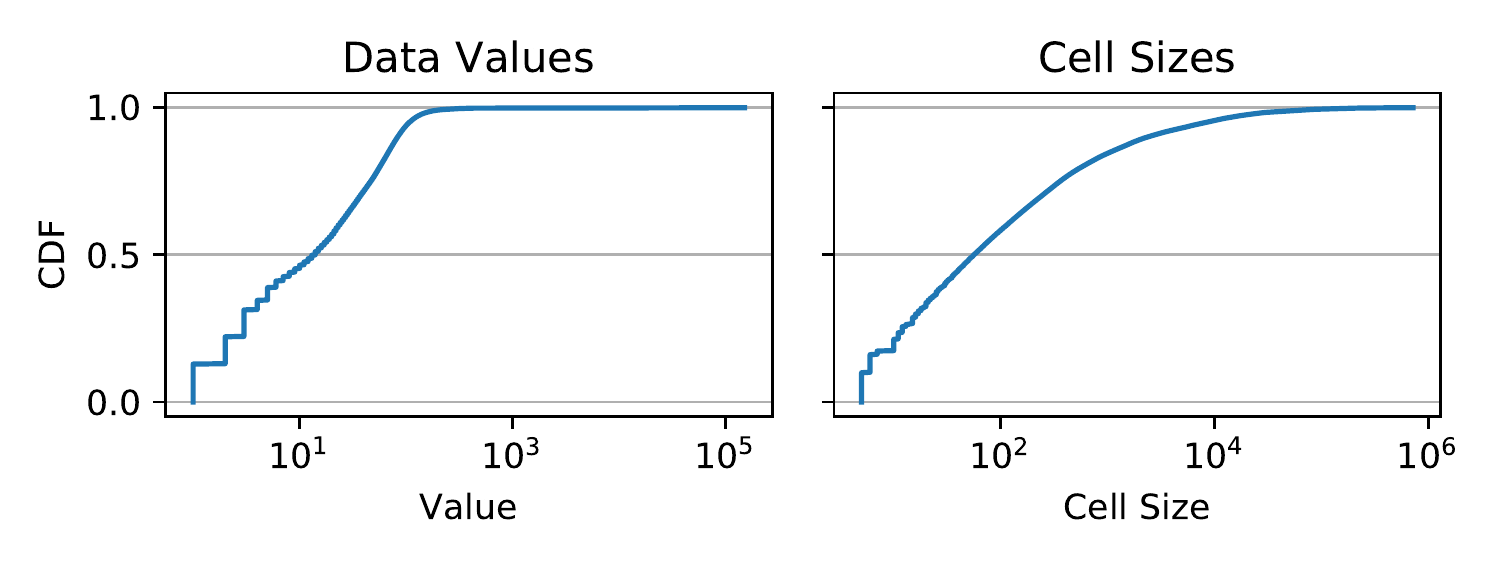}
    \vspace{-2em}
    \caption{Microsoft data values and cell sizes.}
    \label{fig:ariacdf}
\end{figure}

In this section we evaluate merge time and accuracy on a production workload from Microsoft that contains 165 million rows of application telemetry data for an integer-valued performance metric.
We group and pre-aggregate based on four columns that encode information about application version, network type, location, and time, resulting in 400 thousand cells.
Notably, these cells do not correspond to equal sized partitions, but have a minimum size of 5 elements, a maximum size of 722044 elements, and an average size of 2380 elements.
Figure~\ref{fig:ariacdf} illustrates the distribution of integer data values and the cell sizes.

\begin{figure}
    \includegraphics[width=\columnwidth]{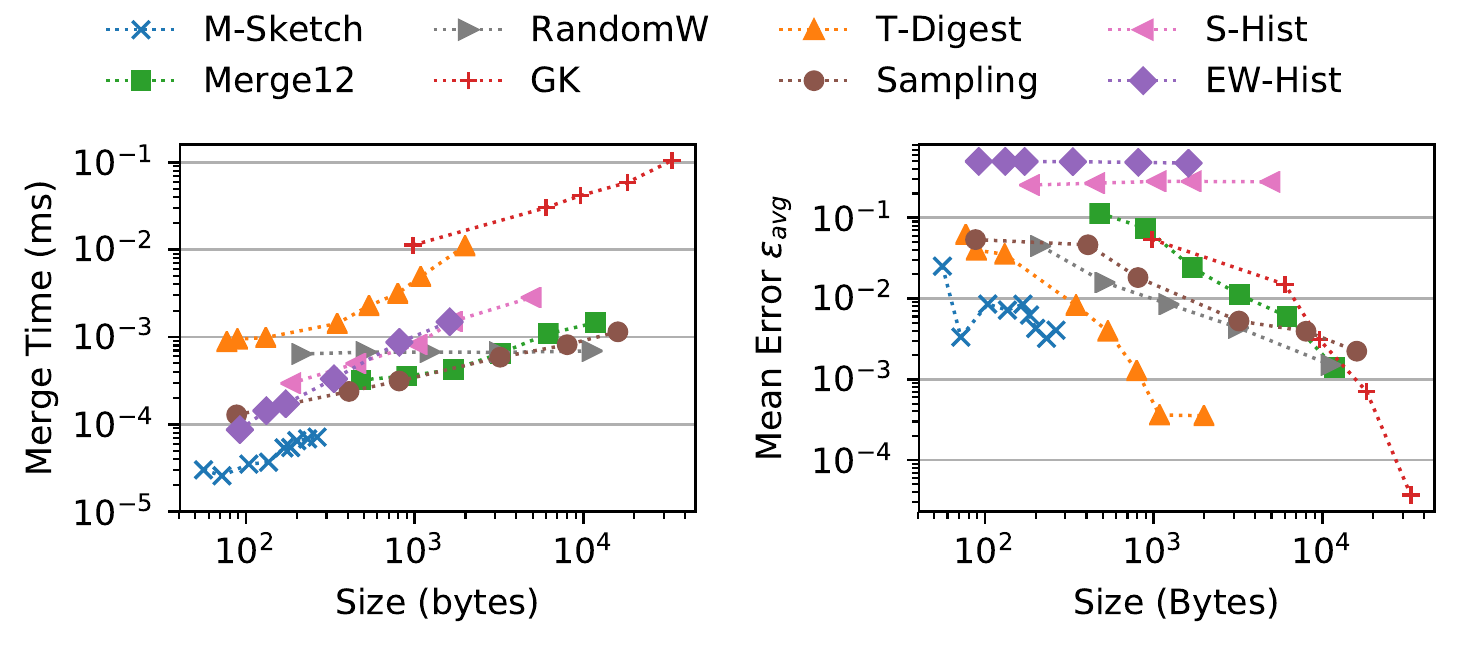}
    \vspace{-2em}
    \caption{Merge times and accuracy on the Microsoft dataset. The merge performance of the \msketch generalizes to workloads with variable sized cells, and exhibits an error rate of $\epsilon_{avg} < .01$.}
    \label{fig:ariamerge}
\end{figure}

Then, we measure the performance and accuracy of merging the cells to perform a quantile aggregation query.
Figure~\ref{fig:ariamerge} illustrates that on this workload with variable sized cells, the \msketch still provides faster merge times than comparably accurate summaries (c.f. Appendix~\ref{sec:appendix_cellsize}).
The \msketch achieves $\epsilon_{avg} < .01$ error when we round estimates to the nearest integer on this integral dataset.
Since the \gk sketch is not strictly mergeable \cite{agarwal2012mergeable}, it grows considerably when merging the heterogenous summaries in this workload to preserve its accuracy.

\section{Error Upper Bounds}
\label{sec:errorboundeval}
\begin{figure}
    \includegraphics[width=\columnwidth]{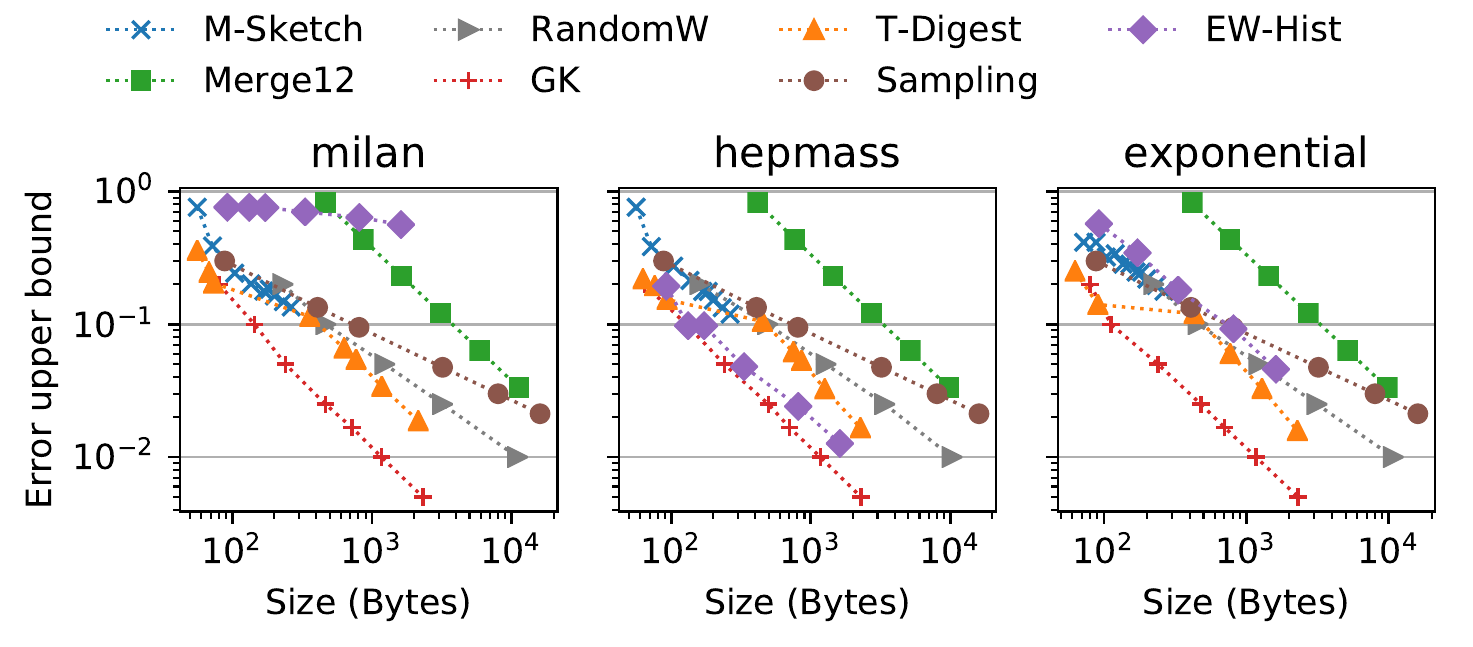}
    \vspace{-2em}
    \caption{Average bound size for summaries of different sizes. No summary is able to provide $\epsilon_{bound} \leq .01$ guarantees with less than 1000 bytes.}
    \vspace{-1em}
    \label{fig:size_bound}
\end{figure}

Thus far we have evaluated observed accuracy.
For comparison, Figure~\ref{fig:size_bound} shows the average guaranteed upper bound error provided by different summaries constructed using pointwise accumulation on the datasets (no merging).
These are in general higher than the observed errors.
We use the \racz routine in Section~\ref{sec:querybounds} to bound the \msketch error.
We omit the \stree since it does not provide upper bounds.
When merging is not a concern, the \gk summary provides the best guaranteed error.

\section{Parallel Merges}
\label{sec:appendix_parallel}
In Section~\ref{sec:evalquerytime}, we evaluated merge time through single-threaded experiments.
We evaluate how well throughput generalizes to parallel aggregation by sharding pre-computed summaries into equal sized batches, and merging the summaries in each batch on an independent worker thread.
After all threads have completed, we combine the merged result from each batch using a sequential merge to obtain a final summary for the complete dataset.

\begin{figure}
    \includegraphics[width=\columnwidth]{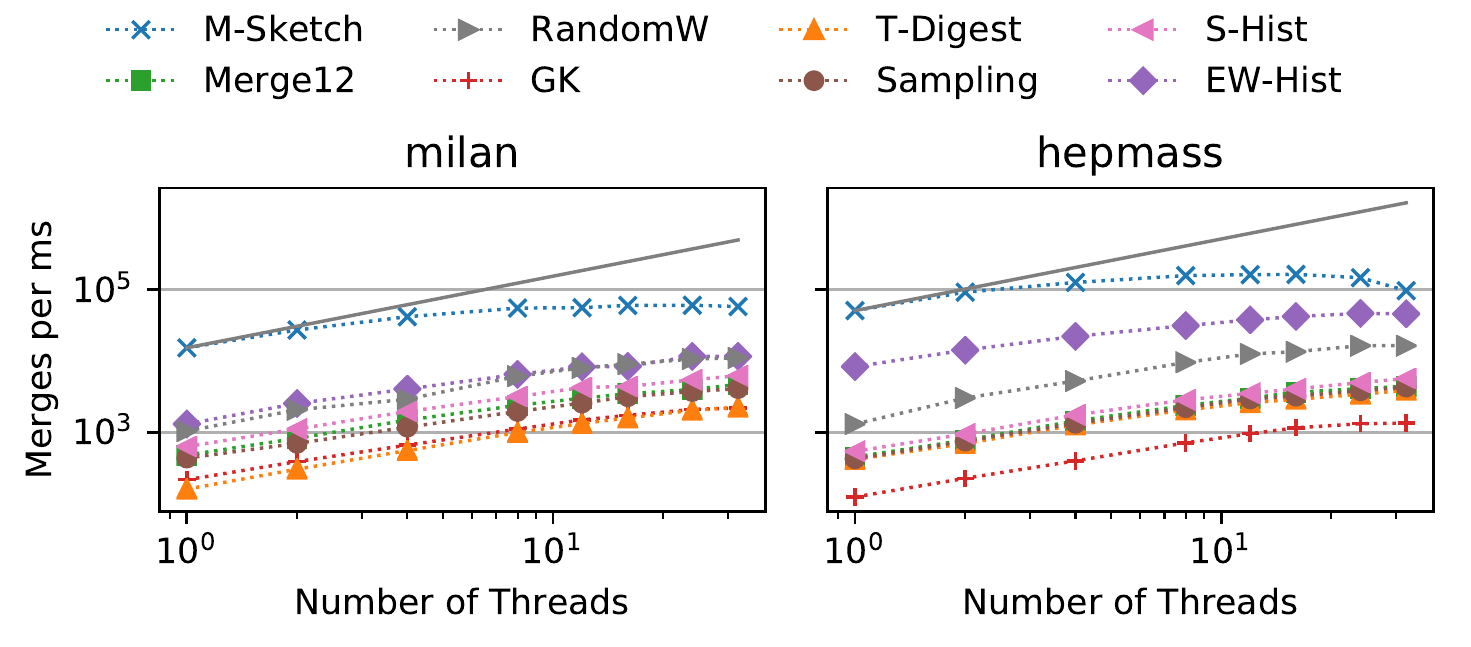}
    \vspace{-2em}
    \caption{Strong scaling of parallel merging. For fixed number of merges, the throughput of the \msketch scales with the number of threads available up to 8-way parallelism, and remains faster than alternatives. The solid line shows ideal \msketch scaling.}
    \label{fig:strongscaling}
\end{figure}

\begin{figure}
    \includegraphics[width=\columnwidth]{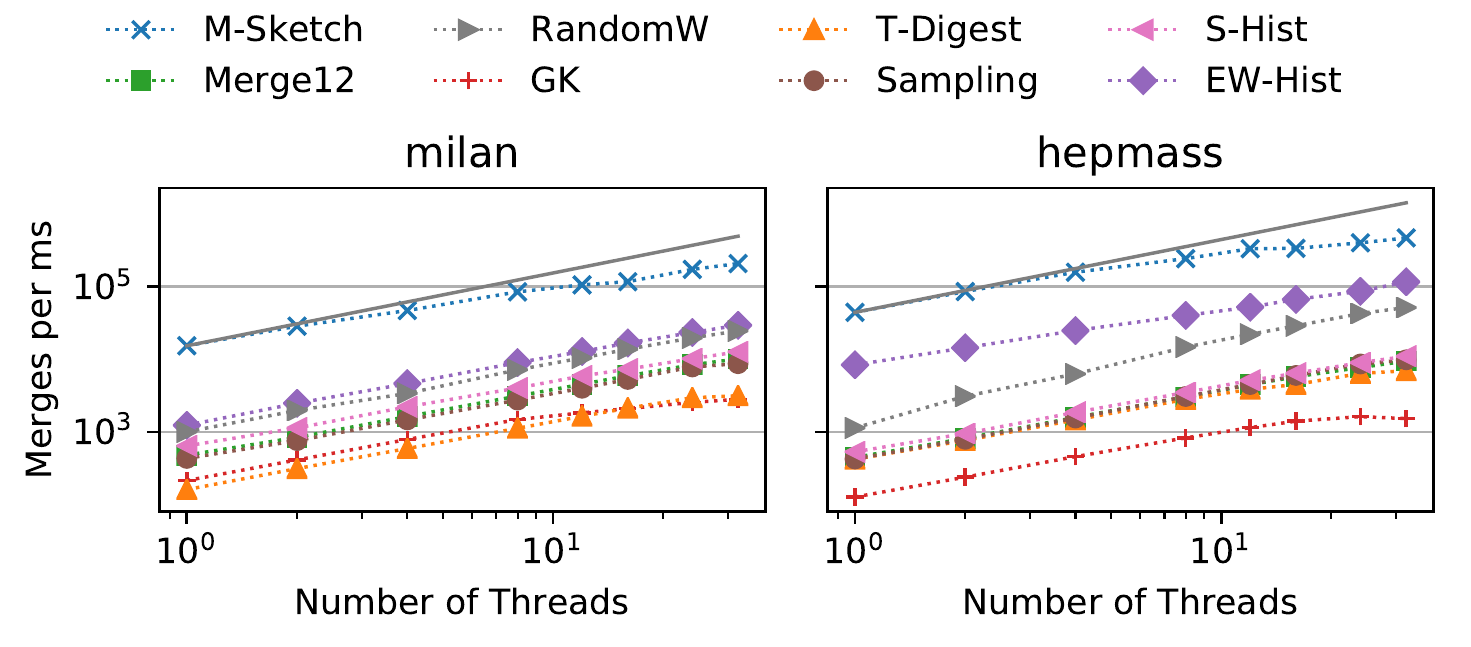}
    \vspace{-2em}
    \caption{Weak scaling of parallel merging. For fixed number of merges per thread, the \msketch and other summaries scale nearly linearly with parallelism.}
    \label{fig:weakscaling}
\end{figure}

In Figure~\ref{fig:strongscaling} we evaluate strong scalability by measuring the total throughput in merging 400 thousand summaries (constructed from blocks of 200 elements) as we increase the number of threads.
In our experiments we needed to duplicate the hepmass dataset to yield 400 thousand summaries, and initialized summaries using the parameters in Table~\ref{tab:sketch_params}.
The \msketch remains faster than alternate summaries as we increase the amount of parallelism, though thread overheads and variance in stragglers limits parallelism on these datasets past 8 threads when there is less work per thread.
In Figure~\ref{fig:weakscaling} we evaluate weak scalability by performing a similar experiment but increase the dataset size alongside thread count, keeping number of merges per thread constant.
Under these conditions, the \msketch and other summaries achieve even better scalability.

These experiments confirm our intuition that since merges can be performed independently, single-threaded performance is indicative of parallel performance, and the relative speedup provided by the \msketch remains stable in parallel settings.
The \msketch and other summaries can be used in more sophisticated distributed aggregation plans as well, such as in~\cite{rusu2012glade,budiu2016sketch}, though since the \msketch is so compact and cheap to merge, multi-level hierarchical aggregation is only profitable when enormous numbers of cores are available.


\end{document}